\documentclass[a4paper,12pt]{article}

\usepackage{amsmath}
\usepackage[psamsfonts]{amssymb}
\usepackage{rsfs}
\usepackage{bm}

\usepackage{cite}
\usepackage[dvips]{graphicx}
\usepackage{color}

\begin{document} 

\begin{titlepage}

\baselineskip 10pt
\hrule 
\vskip 5pt
\leftline{}
\leftline{Chiba Univ./KEK Preprint
          \hfill   \small \hbox{\bf CHIBA-EP-150}}
\leftline{\hfill   \small \hbox{\bf KEK Preprint 2005-6}}
\leftline{\hfill   \small \hbox{hep-ph/0504054}}
\leftline{\hfill   \small \hbox{June 2005}}
\vskip 5pt
\baselineskip 14pt
\hrule 
\vskip 1.0cm
\centerline{\Large\bf 
Numerical evidence for the existence      
} 
\vskip 0.3cm
\centerline{\Large\bf  
of a novel magnetic condensation
}
\vskip 0.3cm
\centerline{\Large\bf  
in Yang-Mills theory
}
\vskip 0.3cm
\centerline{\large\bf  
}

\vskip 0.5cm

\centerline{{\bf 
S. Kato$^{\sharp,{1}}$, 
K.-I. Kondo$^{\dagger,\ddagger,{2}}$,  
T. Murakami$^{\ddagger,{3}}$,
A. Shibata$^{\flat,{4}}$, 
and 
T. Shinohara$^{\ddagger,{5}}$
}}  
\vskip 0.5cm
\centerline{\it
${}^{\sharp}$Takamatsu National College of Technology, Takamatsu 761-8058, Japan
}
\vskip 0.3cm
\centerline{\it
${}^{\dagger}$Department of Physics, Faculty of Science, 
Chiba University, Chiba 263-8522, Japan
}
\vskip 0.3cm
\centerline{\it
${}^{\ddagger}$Graduate School of Science and Technology, 
Chiba University, Chiba 263-8522, Japan
}
\vskip 0.3cm
\centerline{\it
${}^{\flat}$Computing Research Center, High Energy Accelerator Research Organization (KEK),  
}
\centerline{\it
Tsukuba 
305-0801, 
Japan
}
\vskip 1cm

\begin{abstract}
We present a first numerical evidence for the existence of a novel magnetic condensate proposed recently by one of the authors in SU(2) Yang-Mills theory.
In our framework, the spontaneously generated color magnetic field identified with the Savvidy vacuum has the microscopic origin and is a consequence of the intrinsic dynamics of the Yang-Mills theory. 
It strongly suggests the Nielsen--Olesen instability of the Savvidy vacuum  disappears and the  stability is restored without the need of the Copenhagen vacuum. 
The implications to the Skyrme--Faddeev model are also discussed. 
These results are obtained through the first implementation of the Cho--Faddeev--Niemi decomposition of the Yang-Mills field on a lattice.

\end{abstract}


\hrule  
\vskip 0.1cm
${}^1$ 
  E-mail:  {\tt kato@takamatsu-nct.ac.jp}
  
${}^2$ 
  E-mail:  {\tt kondok@faculty.chiba-u.jp}
  
${}^3$ 
  E-mail:  {\tt tom@cuphd.nd.chiba-u.ac.jp}
  
${}^4$ 
  E-mail:  {\tt akihiro.shibata@kek.jp}
  
${}^5$ 
  E-mail:  {\tt sinohara@cuphd.nd.chiba-u.ac.jp}

\par 
\par\noindent


\vskip 0.5cm

\newpage
\pagenumbering{roman}
\tableofcontents




\end{titlepage}


\pagenumbering{arabic}

\baselineskip 14pt
\section{Introduction}
\setcounter{equation}{0}

In the SU(2) Yang-Mills theory, Savvidy\cite{Savvidy77} has discovered according to the renormalization group equation that a non-perturbative vacuum with dynamically generated color magnetic field $H$ has lower vacuum energy density than the perturbative vacuum. 
This is possible only for the non-Abelian gauge theory with asymptotic freedom.  
Immediately after this discovery, however, Nielsen and Olesen \cite{NO78} have pointed out that the effective potential $V(H)$ of the color magnetic field $H$, when calculated explicitly at one-loop level, develops a pure imaginary part;
 The real part of $V(H)$ has an absolute minimum at $H=H_0\not=0$ away from $H=0$ and satisfies the renormalization group equation in agreement with the Savvidy argument, 
while the non-vanishing imaginary part also satisfies the renormalization group equation without the renormalization scale $\mu$ dependence. 
The presence of the pure imaginary part implies that the Savvidy vacuum becomes unstable due to gluon--antigluon pair annihilation. 
Since the energy eigenvalue $E_n$ of the {\it massless} {\it off-diagonal} gluons with   spin $S=1$ and $S_z=\pm 1$,
 in the constant external magnetic field $H_z:=H_{12}$ is given by
\begin{equation}
 E_n^\pm=\sqrt{k_z^2+2gH_z(n+1/2) + 2gH_z S_z} \quad (n=0,1,2,\cdots) ,
\end{equation}
  the Nielsen--Olesen instability is also understood as originating from the tachyon mode $n=0, S_z=-1$, i.e., the lowest Landau level with antiparallel spin to the external magnetic field, 
\begin{equation}
E_0^-= \sqrt{k_z^2-gH_z} .
\end{equation}
In fact, $E_0^-$ becomes pure imaginary in the low-energy region $k_z^2 < gH$.

On the other hand, it is well known that in QED without asymptotic freedom, the non-zero magnetic field does not lower the vacuum energy and hence no magnetic condensation is expected to occur.
Incidentally, external electric field always destabilizes the vacuum  by causing 
electron-positron pair creation in QED 
and 
gluon pair annihilation in Yang-Mills theory. 
Therefore, no spontaneous generation of electric field is expected in both Abelian and non-Abelian gauge theories.

The Nielsen--Olesen instability of the Savvidy vacuum was derived based on the one-loop calculation of the effective potential.  Therefore,  some people consider it as indicating unreliability of the lowest-order loop calculation, i.e., artifact of the approximation.
However, there have been published a huge amount of papers
dealing with the problem of the unstable modes since the Nielsen-Olesen paper, including the
stabilization by higher order terms \cite{Copenhagen}. 
Moreover, the same problem exists also in the supersymmetric Yang-Mills theory in which the higher-order loop corrections are absent, because the covariantly constant background field strength is not supersymmetric \cite{Kay83}. 

A way to circumvent the instability of the Savvidy vacuum is to introduce the magnetic domains with a finite extension into the Yang-Mills vacuum,  in each of which the tachyon mode does not appear as far as  $k_z^2>gH_z$. 
This resolution  is called the Copenhagen vacuum. 
However, the Copenhagen vacuum breaks the Lorentz invariance and color invariance explicitly. 
This issue has been re-examined recently by Cho and his collaborators \cite{Cho03}. 

What type of vacuum is allowed and preferred in the Yang-Mills theory is an important question related to the physical picture of quark confinement.  
Can the instability be resolved even in the one-loop level by a new mechanism? 

First, it is instructive to recall the assumptions taken in Nielsen and Olesen \cite{NO78}. 

\begin{enumerate}
 \item 
The color magnetic field $\vec{H}$ has a uniform magnitude $|\vec{H}|$ in spacetime and  a specific direction $H_{12}=H_z$ (The direction is identified with the quantization axis of the off-diagonal gluon spin).

 \item 
A background gauge is taken as a gauge fixing condition.  
(Note that the background gauge is exactly the same as the {\it Maximal Abelian gauge} \cite{KLSW87} which has been adopted in recent investigations on quark confinement based on  the dual superconductor picture \cite{dualsuper}.) 

\item 
 The off-diagonal gluons are treated as {\it massless} throughout the analysis. 
\end{enumerate}
 
Now we would like to remind you of the facts which have been obtained by the recent investigations on quark confinement since 1990:
\begin{enumerate}
 \item 
In the Maximal Abelian gauge, infrared Abelian dominance \cite{tHooft81,EI82} and magnetic monopole dominance are observed, as first confirmed \cite{SY90} in the numerical simulations on the lattice, see \cite{reviews} for reviews.  
 \item 
The off-diagonal gluons acquire the mass $M$ which is much larger than the diagonal gluon mass \cite{AS99,BCGMP03}.  
The off-diagonal gluon mass $M$ measured on a lattice  is $M \cong 1.2\rm{GeV}$.
See Ref.\cite{KondoI,KondoII,KondoIV,Schaden99,Kondo01,EW03,Dudaletal04} for analytical works.

\end{enumerate}

In the previous work \cite{Kondo04}, the stability of the Savvidy vacuum has been re-examined  by taking into account these facts 
   and a scenario of eliminating  the Nielsen--Olesen instability  has been proposed  to recover the stability of the vacuum: 
A novel type of color magnetic condensation  originating from  magnetic monopoles  can occur and provides the mass of off-diagonal gluons in the  Yang-Mills theory.  
Moreover, a novel magnetic condensation  removes the tachyon mode of the off-diagonal gluon and 
 the Nielsen--Olesen instability of Savvidy vacuum disappears to restore the stability of the magnetic vacuum, if the magnetic condensation is sufficiently large. 

 The dynamical mass generation for the off-diagonal gluons enables us to explain the infrared Abelian dominance and monopole dominance by way of a non-Abelian Stokes theorem.  
These are quite natural and  consistent results for understanding quark confinement, since the condensation of magnetic monopoles is the key concept in the dual superconductor picture. 
Therefore, quark confinement can be compatible with the stability of the Savvidy vacuum without resorting to the Copenhagen vacuum.

The above claims were confirmed at least to one-loop order in the continuum theory by calculating the effective potential \cite{Kondo04}.  
As a technical device, we have applied the Cho--Faddeev--Niemi (CFN)\cite{Cho80,FN98} decomposition  to SU(2) Yang-Mills theory to extract the magnetic monopole degrees of freedom explicitly from the non-Abelian gauge potential.

The purpose of this paper is to go beyond the previous analytical calculations and to confirm some of the above claims by using the numerical simulations on a lattice. 
This paper is organized as follows.
In section 2, we review the CFN decomposition which plays a crucial role in this paper and summarize the analytical results and some predictions obtained in the previous papers \cite{Kondo04}. 
In section 3, we argue how the CFN variables on a lattice are defined to perform the  numerical simulations on a lattice.  
Our definition of the CFN decomposition on a lattice reproduces the   expressions of the continuum formulation from the lattice counterparts, in the naive continuum limit of the lattice spacing going to zero. 
Moreover, we simulate the lattice Yang-Mills theory without breaking the global  SU(2) symmetry respected by the CFN variable. 
This section constitutes  a crucial step to discriminate our approach from the other works which are apparently similar to ours.  
In section 4, we present the first results of numerical simulations based on the lattice gauge theory using the lattice CFN variables set up in the previous section.  
The first numerical evidence is obtained for the existence of two types of vacuum condensates, which supports the recovery of the stability in the Savvidy vacuum as claimed in \cite{Kondo04}. 
The final section is devoted to conclusion and discussion.

In Appendix A, we show how the gauge invariance of the Yang-Mills theory is expressed in terms of the CFN variables.  Then we discuss how the gauge fixing is performed to eliminate the gauge degrees of freedom, especially in the Maximal Abelian gauge. 
In Appendix B, we summarize the relationship between the gauge fixing on a lattice and the continuum limit. It is shown explicitly that 
the gauge fixing procedures on a lattice which is actually used in the numerical simulations reduce to those known in the continuum formulation \cite{Cho80,FN98,Kondo04} in the naive continuum limit. 

\section{Results and predictions from analytical works}
\setcounter{equation}{0}

\subsection{CFN decomposition in the continuum}

%
%

We adopt the Cho-Faddeev-Niemi (CFN) decomposition for the non-Abelian gauge field \cite{Cho80,FN98,Shabanov99,Gies01}: 
By introducing  a unit vector field $\bm{n}(x)$ with three components, i.e., 
$\bm{n}(x) \cdot \bm{n}(x) := n^A(x) n^A(x) = 1$ $(A=1,2,3)$,  
the non-Abelian gauge field $\mathscr{A}_\mu(x)$ in the SU(2) Yang-Mills theory is decomposed as
\begin{align}
   \mathscr{A}_\mu(x) 
=\overbrace{ \underbrace{ c_\mu(x) \bm{n}(x)}_{\mathbb{C}_\mu(x)}
 +    \underbrace{ g^{-1}  \partial_\mu \bm{n}(x)  \times \bm{n}(x) }_{\mathbb{B}_\mu(x)}}^{\mathbb{V}_\mu(x)}  
 + \mathbb{X}_\mu (x) ,
\end{align}
where we have used the notation: 
$\mathbb{C}_\mu(x):=c_\mu(x) \bm{n}(x)$,  
$\mathbb{B}_\mu(x):=g^{-1} \partial_\mu \bm{n}(x) \times \bm{n}(x)$
and
$\mathbb{V}_\mu(x):=\mathbb{C}_\mu(x)+\mathbb{B}_\mu(x)$. 
By definition,  
$\mathbb{C}_\mu(x)$ is parallel to $\bm{n}(x)$, while $\mathbb{B}_\mu(x)$ is orthogonal to $\bm{n}(x)$.  We require $\mathbb{X}_\mu(x)$ to be orthogonal to $\bm{n}(x)$,  i.e., 
$\bm{n}(x) \cdot \mathbb{X}_\mu(x)=0$.  
We call $\mathbb{C}_\mu(x)$ the restricted potential, while $\mathbb{X}_\mu(x)$ is called the gauge-covariant potential and 
 $\mathbb{B}_\mu(x)$ is called the non-Abelian magnetic potential. 
In the naive Abelian projection,  
$\mathbb{C}_\mu(x)$ corresponds to the diagonal component, 
while $\mathbb{X}_\mu(x)$ corresponds to the off-diagonal component, 
apart from the vanishing magnetic part $\mathbb{B}_\mu(x)$.

Accordingly, the non-Abelian field strength $\mathscr{F}_{\mu\nu}(x)$ is decomposed as
\begin{align}
  \mathscr{F}_{\mu\nu} := \partial_\mu \mathscr{A}_\nu - \partial_\nu \mathscr{A}_\mu + g \mathscr{A}_\mu \times \mathscr{A}_\nu
= \mathbb{E}_{\mu\nu} + \mathbb{H}_{\mu\nu} + \hat{D}_\mu \mathbb{X}_\nu - \hat{D}_\nu \mathbb{X}_\mu + g \mathbb{X}_\mu \times \mathbb{X}_\nu , 
\end{align}
where we have introduced the covariant derivative in the background field $\mathbb{V}_\mu$ by 
$
  \hat{D}_\mu[\mathbb{V}]  \equiv \hat{D}_\mu  := \partial_\mu   + g \mathbb{V}_\mu \times  ,
$
and defined the two kinds of field strength:
\begin{align}
  \mathbb{E}_{\mu\nu} =& E_{\mu\nu} \bm{n}, \quad
E_{\mu\nu} := \partial_\mu c_\nu - \partial_\nu c_\mu ,
\\
  \mathbb{H}_{\mu\nu}  
=& \partial_\mu \mathbb{B}_\nu - \partial_\nu \mathbb{B}_\mu + g \mathbb{B}_\mu \times \mathbb{B}_\nu
 .
\end{align}
Due to the special definition of $\mathbb{B}_\mu$,  the magnetic field strength  $\mathbb{H}_{\mu\nu}$ is rewritten as
\begin{align}
  \mathbb{H}_{\mu\nu}  
&=   - g \mathbb{B}_\mu \times \mathbb{B}_\nu 
=  - g^{-1}   (\partial_\mu \bm{n} \times \partial_\nu \bm{n})  
=  H_{\mu\nu} \bm{n}, 
\\
H_{\mu\nu} &:=  - g^{-1} \bm{n} \cdot (\partial_\mu \bm{n} \times \partial_\nu \bm{n})  .
\end{align}
where we have used a fact that $\mathbb{H}_{\mu\nu}$ is parallel to $\bm{n}$. 
Moreover, $H_{\mu\nu}$ is shown to be locally closed and hence it can be exact locally. In other words, we can introduce the Abelian magnetic potential $h_\mu$ for $H_{\mu\nu}$:
\begin{align}
H_{\mu\nu} := - g^{-1} \bm{n} \cdot (\partial_\mu \bm{n} \times \partial_\nu \bm{n})  
= \partial_\mu h_\nu - \partial_\nu h_\mu .
\end{align}
Thus we can introduce two kinds of {\it Abelian} potential $c_\mu$ and $h_\mu$ and the corresponding {\it Abelian} field strength $E_{\mu\nu}=\partial_\mu c_\nu - \partial_\nu c_\mu$ and $H_{\mu\nu}=\partial_\mu h_\nu - \partial_\nu h_\mu$. 
We call $c_\mu$ the (Abelian) electric potential and $h_\mu$ the (Abelian) magnetic potential 
(partial duality), because  $H_{\mu\nu}$ represents the color magnetic field generated by magnetic monopoles \cite{KondoII}. 
The CFN decomposition is useful to extract the topological configurations explicitly, such as a magnetic monopole (of Wu-Yang type), one instanton (of BPST type), and multi-instantons (of Witten type).   
The gauge invariance of the Yang-Mills theory in terms of the CFN variable is discussed in Appendix A.

\subsection{Advantages of our method using the CFN decomposition}

We enumerate some advantages and and  characteristics in our treatment of the magnetic vacuum using the CFN decomposition. (Some of them have already been emphasized by Cho \cite{Cho03}.) 

\begin{enumerate}
 \item 
In our approach using the CFN decomposition, 
the direction  of the color magnetic field 
$\mathbb{H}_{\mu\nu}(x)=H_{\mu\nu}(x) \bm{n}(x)$ 
can be chosen arbitrary at every spacetime  point $x$ by using a unit vector $\bm{n}(x)$ indicating the color direction.
The Lorentz symmetry and color (global gauge) symmetry are not broken by considering 
$
\|\mathbb{H}\| :=   \sqrt{\mathbb{H}_{\mu\nu} \cdot \mathbb{H}_{\mu\nu}} \equiv  g^{-1}   \sqrt{(\bm{n} \cdot (\partial_\mu \bm{n} \times \partial_\nu \bm{n}))^2}
$. 
It is invariant also under the color reflection, 
$\bm{n}(x) \rightarrow - \bm{n}(x)$.

 \item 
This formalism enables us to specify the physical origin of magnetic condensation as arising from the magnetic monopole 
through the relation, 
$
H_{\mu\nu}(x) := - g^{-1} \bm{n}(x) \cdot (\partial_\mu \bm{n}(x) \times \partial_\nu \bm{n}(x))
$.  
This gives a microscopic description of the dynamically generated color magnetic field $H_{\mu\nu}$ which is not necessarily uniform in spacetime, in contrast to the Savvidy, Nielsen and Olesen.

\item 
We can discuss the implications to the Skyrme-Faddeev model\cite{FN97} which is supposed to be a low-energy effective theory of Yang-Mills theory.    
This model is expected to describe  glueballs as knot solitons.

\item
 The non-Abelian Wilson loop operator can be rewritten in terms of the CFN variables through the Diakonov--Petrov version of the non-Abelian Stokes theorem \cite{DP89,KondoIV}.  Hence we can separate the contribution from the magnetic variables in the Wilson loop average to examine the magnetic monopole dominance.

\end{enumerate}

\subsection{Predictions}

In the previous work \cite{Kondo04} the following issues have been discussed. 
\begin{enumerate} 
 \item 
 A novel type of vacuum condensation
$\langle \mathbb{B}_\mu \cdot \mathbb{B}_\mu \rangle>0$
can occur in addition to the magnetic condensation 
$\langle \| \mathbb{H} \| \rangle:=\left< \sqrt{(\mathbb{B}_\mu \times \mathbb{B}_\nu)^2} \right> >0$.   
Here $\langle \mathbb{B}_\mu \cdot \mathbb{B}_\mu \rangle>0$, 
 is called the magnetic condensation of mass dimension two and
 $\langle \| \mathbb{H} \| \rangle>0$ represents the 
spontaneous or dynamical generation of color magnetic field corresponding to the Savvidy vacuum. 
They are caused by gluonic interactions due to  magnetic monopole degrees of freedom which are extracted by the CFN decomposition and are expressed through $\bm{n}$, i.e.,  
\begin{align}
  g^2 \mathbb{B}_\mu \cdot \mathbb{B}_\mu &=  (\partial_\rho \bm{n})^2   ,
  \\
  \|\mathbb{H}\| :=   
  \sqrt{\mathbb{H}_{\mu\nu} \cdot \mathbb{H}_{\mu\nu}}
   \equiv  
\sqrt{(g\mathbb{B}_\mu \times \mathbb{B}_\nu)^2}   
&= g^{-1}   \sqrt{(\partial_\mu \bm{n} \times \partial_\nu \bm{n})^2}   . 
\end{align}

 \item 
  If a novel type of magnetic condensation occurs $\langle \mathbb{B}_\mu \cdot \mathbb{B}_\mu \rangle >0$, then the off-diagonal gluons $\mathbb{X}_\mu$ acquire their mass $M_X$ through the relationship $M_X^2=g^2 \langle \mathbb{B}_\mu \cdot \mathbb{B}_\mu \rangle > 0$. 
Then the infrared Abelian dominance and the magnetic monopole dominance follows immediately from this fact,  supporting the dual superconductor picture for quark confinement.

 \item 
The energy level (spectrum) of the off-diagonal gluons is shifted by $M_X^2$, i.e., 
$E_0^- \rightarrow \sqrt{k^2+M_X^2-gH_0}$. 
 If the off-diagonal gluon mass $M_X$ obtained in this way is sufficiently large so that 
\begin{align}
 M_X^2 \ =g^2 \langle \mathbb{B}_\mu \cdot \mathbb{B}_\mu \rangle > \langle g \|\mathbb{H}\|  \rangle ,
\end{align}
 the tachyon mode is eliminated and the stability of the Savvidy vacuum is restored. 
 Therefore, a criterion of stability restoration is given by 
 \begin{align}
 r  := \frac{M_X^2}{g\|\mathbb{H}\|} = \frac{g^2 \langle \mathbb{B}_\rho \cdot \mathbb{B}_\rho \rangle}{g \langle \|\mathbb{H}\|  \rangle}
 = \frac{\left< (\partial_\rho \bm{n})^2 \right>}{\left< \sqrt{(\partial_\mu \bm{n} \times \partial_\nu \bm{n})^2}\right>}  > 1 .
\end{align}

\end{enumerate}

In fact, the above statements are supported from analytical works as follows. 
Even in the massive case, the existence of a magnetic condensation has been shown 
$
  \langle g \|\mathbb{H}\|  \rangle > 0 ,
$
based on the effective potential in  the one-loop level (improved by the renormalization group)\cite{Kondo04}
where the Maximal Abelian gauge written in terms of the CFN variables,
\begin{align}
 \bm{\chi} := D_\mu[\mathbb{V}] \mathbb{X}_\mu = 0 ,
\end{align}
is adopted. 

Then, the existence of another magnetic condensation,  
$
 g^2 \langle \mathbb{B}_\mu \cdot \mathbb{B}_\mu \rangle  >0 ,
$
can be shown\cite{Kondo04}  based on a simple mathematical identity 
$
 (\mathbb{B}_\mu \times \mathbb{B}_\nu) \cdot (\mathbb{B}_\mu \times \mathbb{B}_\nu)
=  (\mathbb{B}_\mu \cdot \mathbb{B}_\mu)^2 - (\mathbb{B}_\mu \cdot \mathbb{B}_\nu) (\mathbb{B}_\mu \cdot \mathbb{B}_\nu) , 
$
which yields a lower bound on $\mathbb{B}_\mu \cdot \mathbb{B}_\mu$,
$
  (\mathbb{B}_\mu \cdot \mathbb{B}_\mu)^2 - (\mathbb{B}_\mu \times \mathbb{B}_\nu) \cdot (\mathbb{B}_\mu \times \mathbb{B}_\nu) = (\mathbb{B}_\mu \cdot \mathbb{B}_\nu) (\mathbb{B}_\mu \cdot \mathbb{B}_\nu) \ge 0, 
$
i.e., 
$
 g^2 \langle \mathbb{B}_\mu \cdot \mathbb{B}_\mu \rangle \ge \langle g \|\mathbb{H}\|  \rangle  ,
$
leading to a lower bound of the ratio 
\begin{align}
 r    \ge 1 .
\end{align}
Then the tachyon mode is removed.  But the possible zero  mode can not be excluded by this bound. 
A stronger bound is obtained\cite{Kondo04} by using the Faddeev--Niemi variable\cite{FN02}, 
$
 g^2 \langle \mathbb{B}_\mu \cdot \mathbb{B}_\mu \rangle \ge \sqrt{2} \langle g \|\mathbb{H}\|  \rangle ,
$
which yields a better lower bound on the ratio,  
\begin{align}
  r   \ge \sqrt{2} .
\end{align}
This bound is also obtained by another method, see e.g. \cite{Ward98}.
Thus the tachyon mode and the zero mode are removed. 
In fact, the effective potential $V(\|\mathbb{H}\|)$ is real-valued for $r\ge1$. 
In particular, the $r \downarrow 0$ limit reproduces the Nielsen-Olesen pure imaginary part, i.e., instability. 

For the above arguments to work in the rigorous sense, the existence of the  magnetic condensation  
$g \langle \|\mathbb{H}\|  \rangle>0$ 
must be shown in the full non-perturbative level beyond the loop calculation. 
This automatically leads to the existence of a novel magnetic condensation 
$g^2 \langle \mathbb{B}_\mu \cdot \mathbb{B}_\mu \rangle$, if such a vacuum is stable. 
The precise value of the ratio $r$ is not yet determined. 
In fact, there is no theoretical upper  bound on $r$, while the lower bound 
 $r \ge \sqrt{2}$ is known. 
Hence, we  perform Monte Carlo simulations on a lattice to attack these issues.

\section{CFN decomposition on a lattice and $\bm{n}$ field ensemble}
\setcounter{equation}{0}

We denote by the CFN-Yang--Mills theory the Yang-Mills theory written in terms of the CFN variables.
The CFN-Yang--Mills theory has the local gauge symmetry $SU(2)_{local}^{\omega} \times [SU(2)/U(1)]_{local}^{\theta}$ larger than the original Yang-Mills theory,
  since we can rotate the CFN variable $\bm{n}(x)$ by angle $\bm{\theta}^{\perp}(x)$ independently of the gauge transformation parameter $\bm{\omega}(x)$ of  $\mathscr{A}_\mu(x)$, see Appendix A and \cite{KMS05} for more details.  
  In order to fix the whole local gauge symmetry, therefore,  we must impose sufficient number of gauge fixing conditions. 
Recently, it has been clarified \cite{KMS05} how the CFN-Yang--Mills theory can be equivalent to the original Yang-Mills theory after the gauge fixing of the local gauge invariance in the continuum formulation. 
This idea is implemented on a lattice as follows.

Now we discuss how to perform the CFN decomposition  on a lattice and define the unit vector field $\bm{n}_{x}$ to generate the ensemble of $\bm{n}$-fields. 
In the whole of this paper, we restrict the gauge group to SU(2).

\subsection{LLG and new MAG}

First of all, we generate the configurations of SU(2) link variables 
$\{ U_{x,\mu} \}$, 
\begin{align}
 U_{x,\mu}= \exp [ - i\epsilon g \mathscr{A}_\mu(x) ] ,
\end{align}
using the standard Wilson action based on the heat bath method 
\cite{Creutz80} 
where $\epsilon$ is the lattice spacing and $g$ is the coupling constant.%
\footnote{It is possible to adopt different relationships between the link variable and the gauge potential, e.g., 
$
  U_{x,\mu}= \exp [ - i\epsilon g \mathscr{A}_\mu(x+\frac{1}{2}\epsilon \hat{\mu}) ] .
$
However, the difference appears only in higher order terms in the lattice spacing $\epsilon$.  They do not affect our main results and hence the difference is neglected in what follows.} 
We use the continuum notation only for the Lie-algebra valued field variables, e.g., $\mathscr{A}_\mu(x)$.

Next, we introduce the functional,
 
\begin{align}
  F_{LLG}[U; \Omega] = \sum_{x,\mu} {\rm tr}(1-{}^{\Omega}U_{x,\mu}) 
  \rightarrow \frac14 \int d^4x \  [(\mathscr{A}_\mu)^{\Omega}(x)]^2  
  \quad (\epsilon \rightarrow 0) ,
\end{align}
where ${}^\Omega{}U_{x,\mu}$ denotes the gauge-transformed link variable  defined by 
${}^\Omega{}U_{x,\mu} := \Omega_{x} U_{x,\mu} \Omega_{x+\mu}^\dagger$ 
with a gauge group element $\Omega_{x}$ being an SU(2) matrix defined on a site $x$, 
and $(\mathscr{A}_\mu)^{\Omega}(x)$ denotes the gauge-transformed potential defined by 
$(\mathscr{A}_\mu)^{\Omega}(x)=\Omega(x)[\mathscr{A}_\mu(x)+ig^{-1}\partial_\mu]\Omega^\dagger(x)=\mathscr{A}_\mu(x)+D_\mu[\mathscr{A}]\omega(x) + O(\omega^2)$ for $\Omega(x)=e^{ig\omega(x)}$. 
Here the arrow indicates the naive continuum limit $\epsilon \rightarrow 0$ of the lattice spacing $\epsilon$ going to zero, see Appendix B. 
Then we impose the Lorentz-Landau gauge or Lattice Landau gauge (LLG) by minimizing the function 
$F_{LLG}[U; \Omega]$ 
with respect to the gauge transformation $\Omega_{x}$ for the given link configurations $\{ U_{x,\mu} \}$, i.e.,
\begin{align}
  \text{minimizing}_{\Omega} F_{LLG}[U; \Omega]
  \rightarrow 
  \text{minimizing}_{\Omega} \int d^4x \  [(\mathscr{A}_\mu)^{\Omega}(x)]^2  .
\end{align}
In the continuum formulation, this is equivalent to imposing the gauge fixing condition
$\partial_\mu \mathscr{A}_\mu(x)=0$. 
Thus this procedure determines a set of gauge-rotation matrices $\{ \Omega_x \}$.  
Note that the LLG fixes the local gauge symmetry $SU(2)_{local}^{\omega}$, while 
  the LLG leaves the global symmetry SU(2)$_{global}^{\omega}$ intact. 
  See Appendix A. 
  

\begin{figure}[htbp]
\begin{center}
\includegraphics[height=8cm]{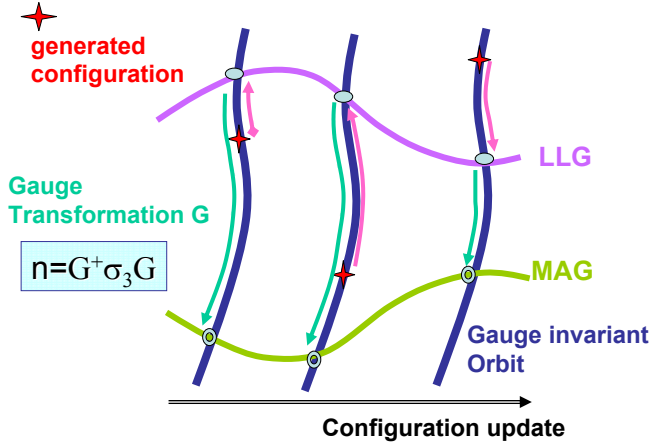}
\caption{\small 
Lattice CFN decomposition obtained by imposing nMAG and LLG. 
}
\label{fig:GF-orbit}
\end{center}
\end{figure}

Subsequently, we impose the new Maximal Abelian gauge%
\footnote{
This procedure is the same as the usual MAG.  However, the meaning is totally different from the usual MAG, as shown in \cite{KMS05}.
} 
 (nMAG) by minimizing the functional 
$F_{MAG}[\tilde{U}; G]$, $\tilde{U}_{x,\mu}:= {}^\Omega{}U_{x,\mu}$ defined by 
\begin{align}
  F_{MAG}[\tilde{U}; G] 
  = \sum_{x,\mu}  {\rm tr}(1-\sigma_3 \ {}^{G}{}\tilde{U}_{x,\mu} \ \sigma_3 \ {}^{G}{}\tilde{U}_{x,\mu}^\dagger ) 
    \rightarrow 
   \int d^4x \  [(A^a_\mu)^{G}(x)]^2  ,
\end{align}
with respect to the gauge transformation $G_{x}$, i.e.,
\begin{align}
  \text{minimizing}_{G} F_{MAG}[\tilde{U}; G] 
    \rightarrow 
  \text{minimizing}_{G} \int d^4x \  [(A^a_\mu)^{G}(x)]^2  ,
\end{align}
where $G(x)=e^{ig\theta(x)}$. 
Here the Cartan decomposition for $\mathscr{A}_\mu(x)$ has been used,
\begin{align}
  \tilde{U}_{x,\mu} = \exp \{ - i \epsilon g \mathscr{A}_\mu(x) \} 
= \exp \{ - i \epsilon g[a_\mu(x) T^3 + A_\mu^a(x) T^a ] \} ,
\end{align}
where $a_\mu$ is called the diagonal gauge field and $A_\mu^a (a=1,2)$ is called the off-diagonal gauge fields with the SU(2) generators $T^A=\sigma^A/2 (A=1,2,3)$.

The nMAG breaks the local gauge symmetry $[SU(2)/U(1)]_{global}^{\theta}$ and  
 leaves the local U(1)$^{\theta}$ symmetry and  the global U(1)$^{\theta}$ symmetry  intact. The superscript $\theta$ indicates that this U(1) is not a subgroup of the SU(2) gauge group for the original Yang-Mills variable $\mathscr{A}_\mu$.  See Appendix A.
Note that the nMAG breaks also the global symmetry $[SU(2)/U(1)]_{global}^{\theta\not=\omega}$, while it does not break $SU(2)_{global}^{\omega=\theta}$. 

The ensemble of $\bm{n}$-fields is constructed as follows. 
See Fig.~\ref{fig:GF-orbit}.
It is shown in Appendix B  that the minimization procedure of the nMAG leads to the construction of $\bm{n}$ according to 
\begin{align}
  \bm{n}_{x} := G_{x}^\dagger \sigma_3 G_{x}
  = n_{x}^A \sigma^A  , 
  \quad \left( n_{x}^A = \frac{1}{2}{\rm tr}[\sigma_A G_{x}^\dagger \sigma_3 G_{x}]  \right) .
\end{align} 
This is because this MAG leads to the gauge fixing for the CFN variables as 
\begin{align}
  F_{MAG}[\tilde{U}; G] 
  = \sum_{x,\mu}  {\rm tr}(1-\bm{n}_{x} \tilde{U}_{x,\mu} \bm{n}_{x+\mu}\tilde{U}_{x,\mu}^\dagger ) 
   := \tilde{F}_{MAG}[\tilde{U}; n] 
  \rightarrow \int d^4x \  [(\mathbb{X}_\mu)(x)]^2  ,
\end{align} 
if we identify the link variable as 
\begin{align}
  \tilde{U}_{x,\mu} = \exp \{ - i \epsilon g[\mathbb{C}_\mu(x) + \mathbb{B}_\mu(x) + \mathbb{X}_\mu(x) ]\} ={}^\Omega{}U_{x,\mu} ,
  \label{LCFN}
\end{align}
which we call the {\it lattice CFN decomposition}. 
Here $\tilde{F}_{MAG}[\tilde{U}; n]$ implies that MAG is also realized as the minimization  with respect to $\bm{n}_x$.
Even if the initial configurations $U_{x,\mu}$ (or $\mathscr{A}_\mu(x)$) are the same, the CFN variables $\mathbb{C}_\mu(x), \mathbb{B}_\mu(x), \mathbb{X}_\mu(x)$ are not necessarily the same if a different $\bm{n}_x$ is adopted. Therefore, 
$\int d^4x \  [(\mathbb{X}_\mu)(x)]^2$ changes the value depending on the choice of $\bm{n}_x$ or $G_{x}$, although it has no explicit dependence on them. 
\footnote{
A different interpretation is as follows. 
The different MAG functional is obtained for the CFN variable as 
\begin{align}
  F_{MAG}[\tilde{U}; G] 
  = \sum_{x,\mu}  {\rm tr}(1-\bm{n}_{x} \tilde{U}_{x,\mu} \bm{n}_{x+\mu}\tilde{U}_{x,\mu}^\dagger )
  := F_{MAG}[\tilde{U}; n] 
  \rightarrow \int d^4x \  [(\mathbb{X}_\mu)^{\Omega}(x)]^2  ,
\end{align} 
if we identify the link variable with the CFN decomposition, 
\begin{align}
  U_{x,\mu} = \exp \{ - i \epsilon g[\mathbb{C}_\mu(x) + \mathbb{B}_\mu(x) + \mathbb{X}_\mu(x) ]\} .
\end{align}
Even in this case, the same gauge fixing condition is obtained,
$D_\mu[\mathbb{V}]\mathbb{X}_\mu=0$, apart from the exceptional case $\omega=\theta$, see Appendix A. 
}

By imposing simultaneously the LLG and the nMAG in this way, we can completely fix the whole local gauge invariance 
$SU(2)_{local}^{\omega} \times [SU(2)/U(1)]_{local}^{\theta}$
of the lattice CFN-Yang--Mills theory. 
The global symmetry $SU(2)_{global}^{\omega=\theta}$ is unbroken.

Here we distinguish two cases related to the global symmetry $SU(2)_{global}$. 

\subsubsection{$SU(2)_{global}$-breaking case}

If the numerical simulations are performed in such a way that  LLG and MAG are close to each other \cite{DHW02},
in the sense that the matrices $G$ connecting LLG and MAG are on average close to the unit ones,  i.e., 
$G_x^A \cong 0$ $(A=1,2,3)$, i.e., $G_x \cong G_x^0 I$, 
for the parameterization of $SU(2)$ matrices, 
\begin{align}
  G_x = G_x^0 I + i G_x^A \sigma^A, \quad
   G_x^0, G_x^A \in \mathbb{R}, \quad
   \sum_{\mu=0}^{3} (G_x^\mu)^2 = 1 .
\end{align}
then we observe  that 
$\bm{n}_x \cong \sigma^3$ or $n_x^A \cong (0,0,1)$, namely,  
 $\bm{n}_x$ are  aligned in the positive 3-direction and hence the non-vanishing vacuum expectation value is observed as 
\begin{align}
  \left< n^A_x \right> = M \delta^{A3} . 
\end{align}
This implies that the global SU(2) symmetry is broken explicitly to a global U(1),
 $SU(2)_{\text{global}} \rightarrow U(1)_{\text{global}}$. 
In the two-point correlation functions, the exponential decay is observed for 
the parallel propagator  
\begin{align}
  \left< n_x^3 n_0^3 \right> \sim \left< n_0^3 \right>\left< n_0^3 \right> + c e^{-m|x|} = M^2 + c e^{-m|x|} , 
\end{align}
and for the perpendicular propagator
\begin{align}
  \frac{1}{2} \sum_{a=1}^{2} \left< n_x^a n_0^a \right> \sim  c' e^{-m'|x|} ,
\end{align}
but $m$ and $m'$ are slightly different, but nearly equal to $0.9 \rm{GeV}$. 
This result was reported by \cite{DHW02} and confirmed also by our preliminary simulations \cite{DSB04}.

\subsubsection{$SU(2)_{global}$-invariant case}

Our main numerical simulations are performed as follows. 
In the continuum formulation,  the CFN variables were introduced as a change of variables which does not break the global gauge symmetry  $SU(2)_{global}$  or "color symmetry", which has a correspondence with the local  gauge symmetry $SU(2)_{local}$ in the original Yang-Mills theory. 
Hence the nMAG can be imposed in terms of the CFN variables without breaking the color symmetry. 
This is a crucial difference between the nMAG based on the CFN decomposition and the conventional MAG based on the ordinary Cartan decomposition which breaks the  $SU(2)_{global}$ explicitly. 
See Appendix A.
Therefore, we must perform the numerical simulations so as to preserve the color symmetry as much as possible.%
\footnote{ 
Whether the color symmetry is spontaneously broken or not is another issue to be investigated separately. 
}
This is in fact possible as follows.

Remember that the MAG on a lattice is achieved by repeatedly performing the gauge transformations. 
In order to preserve the global SU(2) symmetry,  
we adopt a random gauge transformation only in the first sweep among the whole sweeps of gauge transformations in the standard iterative gauge fixing procedure for the MAG.
This procedure moves an ensemble of unit vectors $\bm{n}_{x}$ to a random ensemble of $\bm{n}_{x}$ which is far away from $\bm{n}_{x}=(0,0,1)$, although this procedure might increase the functional $F_{MAG}$. 
Then we search for the local minima around this configuration of $\bm{n}_{x}$ by performing the successive gauge transformations. 
The first random gauge transformation as well as the subsequent gauge transformations are accumulated to obtain the gauge transformation matrix $G$ by which $\bm{n}$ is constructed.
Beginning with the LLG and ending with the MAG in this way, we can impose both LLG and MAG simultaneously.

Our numerical simulations are performed by using the standard Wilson action and periodic boundary conditions under the following conditions. 
After the thermalization of 3000 sweeps starting with cold initial condition, we have obtained 50 samples of configurations at 100 sweep intervals. 
For LLG and MAG, we have used the over relaxation algorithm.

The data of numerical simulations in Table~\ref{table:vevn} show the vanishing vacuum expectation value  
\begin{align}
  \left< n^A_x \right> = 0  \quad (A=1,2,3) . 
\end{align}

\begin{table}[h]
\caption{The magnetization $<n^A_x>$ on the $16^4$ lattice at $\beta=2.4$.}
\label{magnet-lma6}
\begin{center}
\begin{tabular}{cll}\hline
      & Mean value & Jack knife error(JKbin=2) \\ \hline
$<n^1>$ & -0.0069695  & $\pm$ 0.010294   \\
$<n^2>$ & 0.011511 & $\pm$ 0.015366 \\
$<n^3>$ & 0.0014141 & $\pm$ 0.013791 \\ \hline
\end{tabular}
\label{table:vevn}
\end{center}
\end{table}

\begin{figure}[htbp]
\begin{center}
\includegraphics[height=5.5cm]{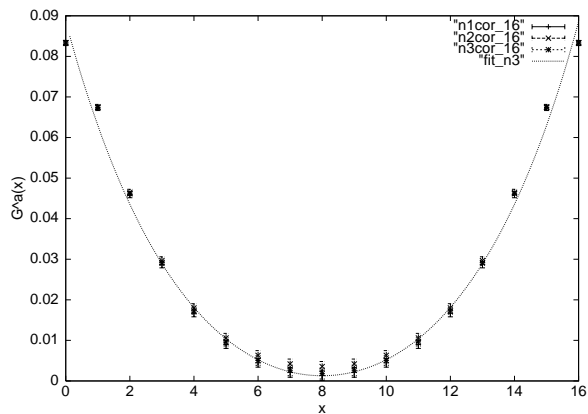}
\caption{\small 
The plots of two-point correlation functions 
$\left< n_x^A n_0^A \right>$ for $A=1,2,3$
along the lattice axis on the $16^4$ lattice at $\beta=2.4$.}
\label{fig:2ptf}
\end{center}
\end{figure}

Moreover, we have measured the two-point correlation functions defined by $G^A(x) := $ $\left< n^A_x n^A_0 \right>$ (no summation over $A$).
The two-point correlation functions exhibit almost the same behavior in all the directions ($A=1,2,3$), see Fig.~\ref{fig:2ptf}. 


These results indicate that {\it the global SU(2) symmetry (color symmetry) is   unbroken in our main simulations}, in contrast to \cite{DHW02}.  This is a crucial point. 

\begin{table}[h]
\caption{Fitted parameters on the $16^4$ lattice at $\beta=2.4$.}
\label{magnet-lma6}
\begin{center}
\begin{tabular}{clll}\hline
      & $a(1)$ & $a(2)$ & $a(3)$ \\ \hline
$G^1$ & 0.0254  & 0.2719 &   -0.2462 \\
$G^2$ & 0.0241 & 0.2745  &  -0.0213 \\
$G^3$ & 0.0284 & 0.2604 &  -0.0271 \\ \hline
\end{tabular}
\label{table:propagator}
\end{center}
\end{table}

The data suggest the exponential decay, 
\begin{align}
  G^A(x) \equiv \left< n_x^A n_0^A \right> \sim  c e^{-m|x|}   \quad (A=1,2,3) . 
\end{align}
The exponential decay implies that there exists the mass gap in the theory. 
This is confirmed as follows. 
In Table~\ref{table:propagator},  we have given the values of three fitting parameters $a(1), a(2), a(3)$  when  $G^3(x)$ is fitted to the cosh-function: $G^3(x)=a(1)*\cosh(a(2)*(x-8))+a(3)$.
Here $a(2)$ corresponds to the mass gap. Since the physical scale $\epsilon(\beta=2.4)$ is 0.26784 in unit of the string tension $\sigma$, the mass gap reads $m_{phys}=a(2)/\epsilon(\beta)= 0.27/(0.26784\sqrt{\sigma}^{-1/2})=1.008\times 440{\rm MeV} = 440 {\rm MeV}$. 
This should be compared with the SU(2) mass gap, $M \cong 1.5 \rm{GeV}$, which could be regarded as the lowest glueball mass \cite{Teper}.

\subsection{Imposing LLG as preconditioning before MAG}

Finally, we explain why the LLG is imposed before taking the MAG. 
From the beginning, we could have imposed the MAG by minimizing the functional,   
\begin{align}
  F_{MAG}[U; G'] 
  = \sum_{x,\mu} {\rm tr}(1-\sigma_3 \ {}^{G'}{}U_{x,\mu} \ \sigma_3 \ {}^{G'}{}U_{x,\mu}^\dagger ) 
    \rightarrow 
   \frac14 \int d^4x \  [(A^a_\mu)^{G'}(x)]^2  ,
\end{align}
with respect to the gauge transformation $G'_{x}$, 
once the link variable configurations 
$\{ U_{x,\mu} \}$, 
$U_{x,\mu}= \exp [ - i \epsilon g \mathscr{A}_\mu(x) ]$ 
are generated using the Wilson action based on the heat bath method.
This is equivalent to minimizing $\tilde{F}_{MAG}[U; n]$ with respect to $\bm{n}_{x}$:
\begin{align}
  F_{MAG}[U; G'] 
  = \sum_{x,\mu}  {\rm tr}(1-\bm{n}_{x} U_{x,\mu} \bm{n}_{x+\mu}U_{x,\mu}^\dagger )
  := \tilde{F}_{MAG}[U; n] 
  \rightarrow \int d^4x \  [(\mathbb{X}_\mu)(x)]^2  ,
\end{align} 
where  the following identifications are made:
\begin{align}
  \bm{n}_{x} := G'{}_{x}^\dagger \sigma_3 G'_{x}
  = n_{x}^A \sigma^A  , 
  \quad \left( n_{x}^A = \frac{1}{2}{\rm tr}[\sigma_A G'{}_{x}^\dagger \sigma_3 G'{}_{x}]  \right) ,
\end{align} 
and 
\begin{align}
  U_{x,\mu} = \exp \{ - i \epsilon g[\mathbb{C}_\mu(x) + \mathbb{B}_\mu(x) + \mathbb{X}_\mu(x) ]\} .
\end{align}

However, it is observed that the resulting ensemble of $\bm{n}_{x}$ becomes random as characterized by the specific two-point correlation function
\begin{align}
  \left< n_x^A n_y^B \right> = \frac{1}{3} \delta^{AB} \delta_{x,y} , \
  \text{i.e.,} \
  = 0 \ (x \not= y) \ \text{or} \ \frac{1}{3}\delta^{AB} \ (x=y) ,
\end{align}
and the vanishing vacuum expectation value
\begin{align}
  \left< n_x^A  \right> = 0 .
\end{align}
There is no correlation among the field $\bm{n}_x$ on the different sites. 
This is because the original link variables $\{ U_{x,\mu} \}$ are generated due to the gauge invariant original action and are distributed   randomly along their gauge orbits. Therefore, the transformation matrix $G'_{x}$ becomes random in  bringing the original gauge field configurations to the gauge fixing hypersurface.  See Fig.~\ref{fig:GF-orbit}.

From the technical viewpoint, this difficulty is avoided if we begin with the ordered link variables $\{ U_{x,\mu} \}$ by a preconditioning which eliminates the randomness.
From this viewpoint, the LLG could be regarded as a preconditioning \cite{DHW02,IS00}.
As we explained in the above, however, the LLG in our approach plays a more  essential and a totally different role of specifying the CFN decomposition by combining LLG with the MAG, rather than merely removing the randomness, as emphasized in \cite{KMS05}.

\subsection{Discriminating our approach from the others}

Although the technique of constructing  the unit vector field $\bm{n}_x$ given above has already  appeared, e.g., in \cite{DHW02,Shabanov01,IS00}, there is a crucial difference between our approach and others.  
In \cite{DHW02,Shabanov01}, the unit vector field $\bm{n}_x$ was regarded as the field variable of the Skyrme--Faddeev model which is conjectured to be a low-energy effective theory of Yang-Mills theory.  However, the precise relationship between the Skyrme--Faddeev model and the original Yang-Mills theory is still under debate.   (The paper \cite{DHW02} concluded with the negative answer.) 
In contrast, our approach can identify the lattice field $\bm{n}_x$ as a lattice version of the CFN field variable $\bm{n}(x)$ obtained by the CFN decomposition of the original gauge potential $\mathscr{A}_\mu(x)$ in Yang-Mills theory. 
In fact, the naive continuum limit of the MAG on the lattice agrees with the MAG for the CFN variable \cite{Kondo04}, see Appendix B.
To the best of our knowledge, such an explicit relationship has not been elucidated in the previous works including \cite{DHW02,Shabanov01,IS00}.  
We do not assume any model written in terms of the unit vector field $\bm{n}_x$, which is regarded as an effective theory of Yang-Mills theory. 

In \cite{DHW02}, it is studied whether the identification of the Skyrme--Faddeev (or Faddeev-Niemi) model as a low-energy effective theory of Yang-Mills theory is efficient or not. 
The Skyrme--Faddeev model can have the same pattern of spontaneous symmetry breaking 
$SU(2) \rightarrow U(1)$ as the nonlinear sigma model.
Therefore, if such spontaneous breaking of the global SU(2) symmetry occurs, two massless Nambu-Goldstone bosons appear and the mass gap disappears.
This is because in the Skyrme--Faddeev model  there are no gauge fields into which the massless Nambu-Goldstone bosons are absorbed through the Higgs mechanism. 
To avoid this unpleasant situation, the global SU(2) symmetry was explicitly broken in \cite{DHW02} by choosing the configuration in the neighborhood of 
$\bm{n}=(0,0,1)$ among a large number of local minima.
This viewpoint is consistent with adopting the ensemble of $\bm{n}_{x}$ aligned in a specific direction, since the $\bm{n}_{x}$ in \cite{DHW02} is the field variable of describing the Skyrme--Faddeev model, which is not necessarily the CFN variable $\bm{n}_{x}$. 
On the contrary, the variable $\bm{n}_{x}$ in our approach always denotes the CFN variable of the original Yang-Mills gauge field, without referring to the Skyrme--Faddeev model.  This viewpoint does not lead to the immediate contradiction. 
The relationship of the Skyrme--Faddeev model and the Yang-Mills theory is discussed in the final section in our framework.

\section{Numerical results: magnetic condensations}
\setcounter{equation}{0}

 We present the first numerical evidence for the existence of two vacuum condensates $\langle g\|\mathbb{H}\| \rangle$ and  $\left<g^2\mathbb B_\mu \cdot \mathbb B_\mu \right>$, indicating   the recovery of stability in the Savvidy vacuum. 
 
\subsection{Setting up the simulations}

We define the lattice derivative \cite{Shabanov01} by 
\begin{align}
  \Delta_\mu \bm{n}(x) := \bm{n}(x+\mu) - \gamma(x) \bm{n}(x) := \bm{s}_\mu(x) ,
\end{align}
which guarantees automatically the orthogonality condition 
$\bm{n}(x) \cdot \Delta_\mu \bm{n}(x)=\bm{n}(x) \cdot \bm{s}(x)=0$ 
on a lattice by choosing $\gamma(x)$ as
\begin{align}
  \gamma(x) = \bm{n}(x) \cdot \bm{n}(x+\mu) .
\end{align}
This is not the case for the naive lattice derivative 
$\partial_\mu^L \bm{n}(x) := \bm{n}(x+\mu) - \bm{n}(x)$. 
Then $\mathbb{B}_\mu(x)$ on a lattice is defined by 
\begin{align}
  g \mathbb{B}_\mu(x) := \bm{s}_\mu(x) \times \bm{n}(x) 
  =  \Delta_\mu \bm{n}(x)  \times \bm{n}(x)
  = \bm{n}(x+\mu) \times \bm{n}(x) .
\end{align}
The squared $g^2 \mathbb{B}_\mu^2(x)$ agrees with $\bm{s}_\mu(x)^2$ just as in the continuum case:
\begin{align}
  g^2 \mathbb{B}_\mu^2(x) = (\bm{s}_\mu(x) \times \bm{n}(x))^2
  = 1 - (\gamma_\mu(x))^2 
  = \bm{s}_\mu(x)^2 = (\Delta_\mu \bm{n}(x))^2 .
\end{align}
A simple calculation shows that 
\begin{align}
  \bm{n}(x) \times (\bm{s}_\mu(x) \times \bm{s}_\nu(x)) 
  = 0 .
\end{align}
This implies that $\bm{s}_\mu(x) \times \bm{s}_\nu(x)$ is parallel to $\bm{n}(x)$ and does not have the components perpendicular to $\bm{n}(x)$. 
Therefore, it is natural to define $g\mathbb{H}_{\mu\nu}(x)$ and 
$gH_{\mu\nu}(x)$ on a lattice by 
\begin{align}
  g \mathbb{H}_{\mu\nu}(x) 
  := -  (\bm{s}_\mu(x) \times \bm{s}_\nu(x)) 
  = g H_{\mu\nu}(x) \bm{n}(x).
\end{align}
and 
\begin{align}
  g H_{\mu\nu}(x) = g \bm{n}(x) \cdot \mathbb{H}_{\mu\nu}(x) 
  := - \bm{n}(x) \cdot (\bm{s}_\mu(x) \times \bm{s}_\nu(x)) 
.
\end{align}
This implies the equality of the squared quantities:
\begin{align}
  g^2 \mathbb{H}_{\mu\nu}^2(x) 
  = (\bm{s}_\mu(x) \times \bm{s}_\nu(x))^2
  = [\bm{n}(x) \cdot (\bm{s}_\mu(x) \times \bm{s}_\nu(x))]^2
  = g^2 H_{\mu\nu}^2(x) .
\end{align}

Our numerical simulations are performed on the lattice  with the   
lattice size $12^{4}$, $24^{4}$, $36^{4}$
by using the standard Wilson action  for the gauge coupling $\beta$ $=2.1\sim2.7$ and periodic boundary conditions. 
Staring with cold initial condition and thermalizing 50*100
sweeps, we have obtained  200 configurations (samples) for $12^4, 36^4$ lattice
and 500 samples for  $24^4$ lattice at intervals of 100 sweeps.
For LLG and MAG, we have used the over relaxation algorithm.




\subsection{Savvidy-like magnetic condensation}


\begin{figure}[htbp]
\begin{center}
\includegraphics[height=5.5cm]{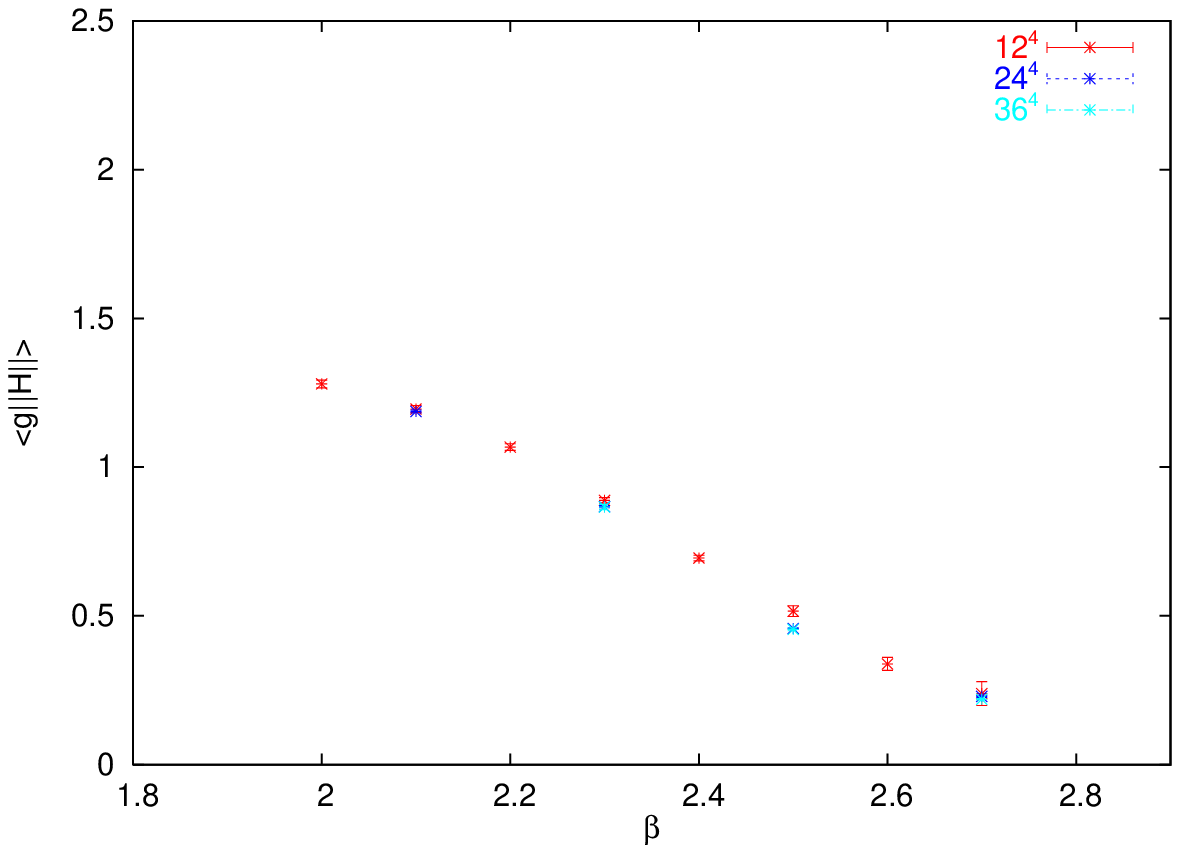}
\caption{\small  The Savvidy-like magnetic condensation:
$\langle g\|\mathbb{H}\| \rangle
:= \langle g\sqrt{\mathbb{H}_{\mu\nu} \cdot \mathbb{H}_{\mu\nu}} \rangle
= \left< \sqrt{(\Delta_\mu \bm{n} \times \Delta_\nu \bm{n})^2} \right>$  
versus $\beta=2.0 \sim 2.7$ on $12^4, 24^4, 36^4$ lattices. 
}
\label{fig:H}
\end{center}
\end{figure}


We have measured the magnetic condensation 
$$
\langle g\|\mathbb{H}(x)\| \rangle
:= \langle g\sqrt{\mathbb{H}_{\mu\nu}(x) \cdot \mathbb{H}_{\mu\nu}(x)} \rangle
= \left< \sqrt{(\Delta_\mu \bm{n}(x) \times \Delta_\nu \bm{n}(x))^2} \right>.
$$
by changing $\beta$ on the lattices with different sizes. 
This corresponds to the Savvidy-like magnetic condensation, but it has microscopic origin written in terms of the field $\bm{n}(x)$ as a part of the gauge potential $\mathscr{A}_\mu$. 
See Fig.~\ref{fig:H} for the numerical values of the dimensionless magnetic condensation versus $\beta=2.0 \sim 2.7$ on $12^4, 24^4, 36^4$ lattices.

\begin{figure}[htbp]
\begin{center}
\includegraphics[height=5.5cm]{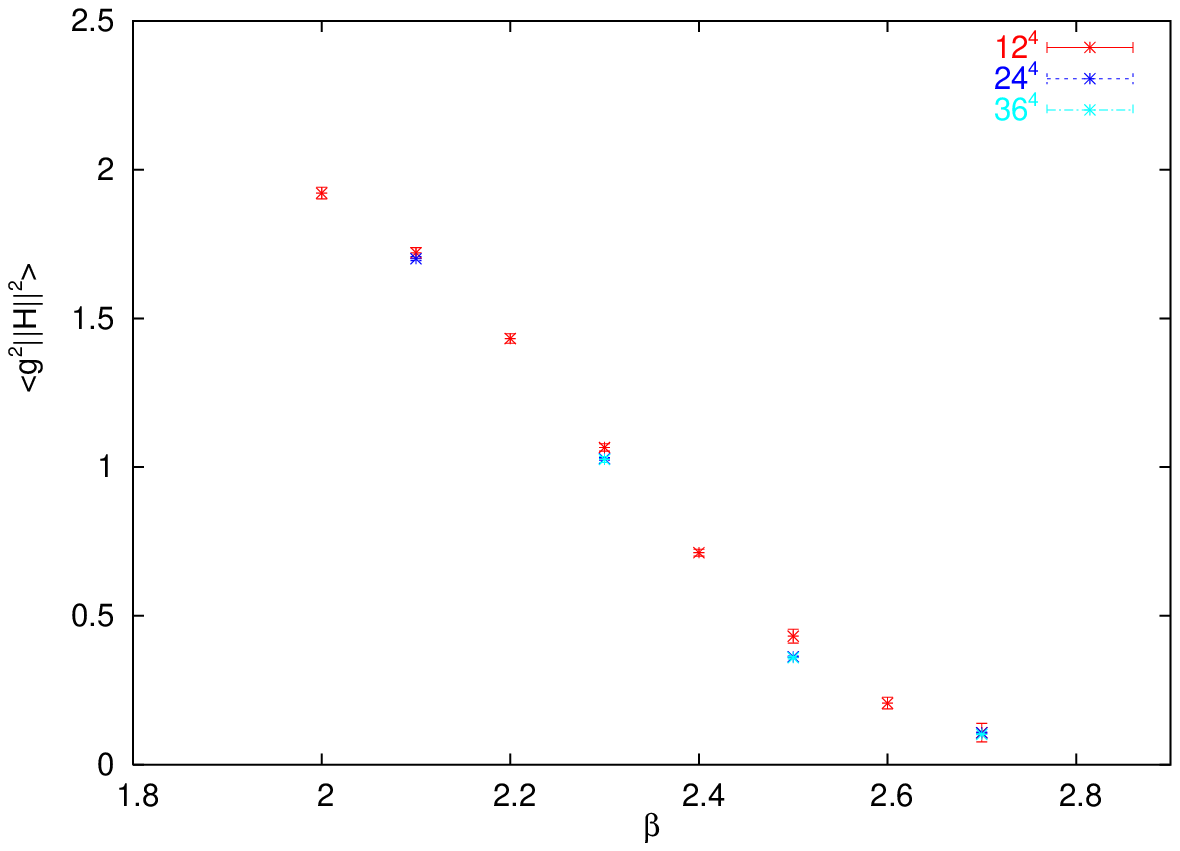}
\caption{\small The squared Savvidy-like magnetic condensation:
$\langle g^2\|\mathbb{H}\|^2 \rangle
:= \langle g^2 \mathbb{H}_{\mu\nu} \cdot \mathbb{H}_{\mu\nu}  \rangle
= \langle (\Delta_\mu \bm{n} \times \Delta_\nu \bm{n})^2 \rangle$ 
versus $\beta=2.0 \sim 2.7$ on $12^4, 24^4, 36^4$ lattices. 
}
\label{fig:HH}
\end{center}
\end{figure}

We have also measured the squared magnetic condensation 
$$
\langle g^2\|\mathbb{H}(x)\|^2 \rangle
:= \langle g^2\mathbb{H}_{\mu\nu}(x) \cdot \mathbb{H}_{\mu\nu}(x) \rangle
= \left< (\Delta_\mu \bm{n}(x) \times \Delta_\nu \bm{n}(x))^2 \right> .
$$
See Fig.~\ref{fig:HH}.
We can estimate the variance,
$\langle (\|\mathbb{H}(x)\|-\langle \|\mathbb{H}(x)\| \rangle)^2 \rangle
= \langle \|\mathbb{H}(x)\|^2 \rangle - \langle\|\mathbb{H}(x)\| \rangle^2$,
and the standard deviation 
$\sigma:=\sqrt{\langle (\|\mathbb{H}(x)\|-\langle \|\mathbb{H}(x)\| \rangle)^2 \rangle}$, as discussed in the effective potential. 


\subsection{A novel magnetic condensation}


\begin{figure}[htbp]
\begin{center}
\includegraphics[height=5.5cm]{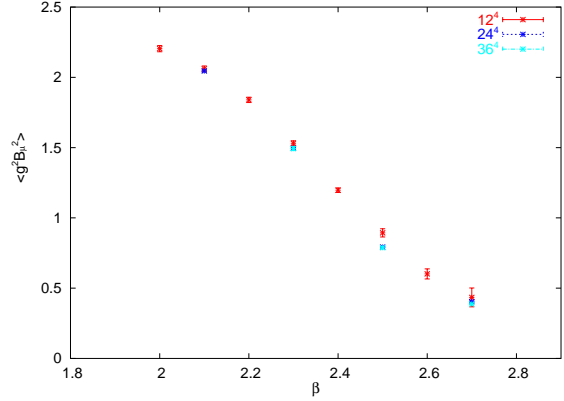}
\caption{\small A novel magnetic condensation  
$\langle g^2 \mathbb{B}_\mu \cdot \mathbb{B}_\mu \rangle
:= \langle (\Delta_\rho \bm{n})^2 \rangle$
versus $\beta=2.0 \sim 2.7$ on $12^4, 24^4, 36^4$ lattices. 
}
\label{fig:BB}
\end{center}
\end{figure}




\begin{figure}[htbp]
\begin{center}
\includegraphics[height=5.0cm]{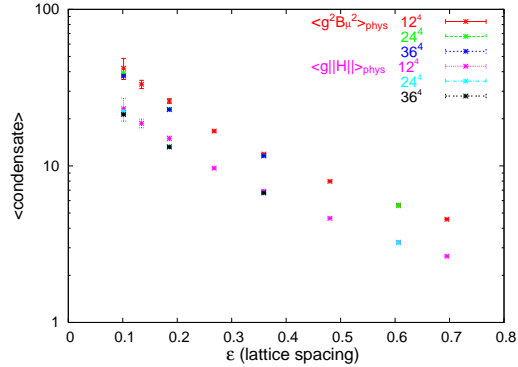}
\caption{\small Two vacuum condensates in units of the physical string tension $\sigma_{phys}$ 
versus the lattice spacing $\epsilon$ on $12^4, 24^4, 36^4$ lattices. 
}
\label{fig:condensations}
\end{center}
\end{figure}



A novel magnetic condensation predicted in \cite{Kondo04} 
$$
\langle g^2 \mathbb{B}_\mu(x) \cdot \mathbb{B}_\mu(x) \rangle
= \left< (\Delta_\rho \bm{n}(x))^2 \right> .
$$
has been measured as shown in Fig.~\ref{fig:BB}.
The value is larger than the Savvidy-like magnetic condensation, as suggested by the analytical lower bound mentioned before.

\begin{figure}[htbp]
\begin{center}
\includegraphics[height=5.5cm]{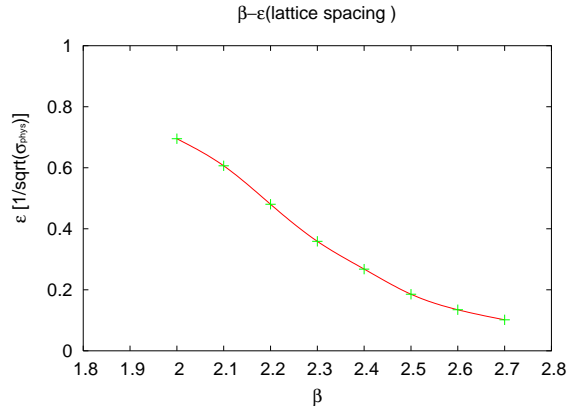}
\caption{\small 
Lattice spacing $\epsilon$ versus $\beta$ (inverse gauge coupling) in units of the physical string tension $\sigma_{phys}$, reproduced from \cite{KKNS98}.}
\label{fig:a-beta}
\end{center}
\end{figure}

All data of two magnetic condensations are collected in Fig.~\ref{fig:condensations} where they are measured in units of the string tension by way of the lattice spacing
$
 \epsilon(\beta) = \sqrt{\sigma(\beta)/\sigma_{\rm phys}} , 
$
as a function $\beta$ (Fig.~\ref{fig:a-beta})
where 
$\sigma_{\rm phys}=(440 \rm{MeV})^2$ is the physical string tension
and $\sigma(\beta)$ is the (dimensionless) lattice string tension (determined by the magnetic monopole part of the Abelian Wilson loop),  
 see \cite{KKNS98} for details.  
 
 The numerical value $M(\beta)$ measured on the lattice for the quantity of mass dimension one is translated into the physical value $M_{phys}$ 
through the relation
$M(\beta)=M_{phys} \cdot \epsilon(\beta)$,
\begin{align}
  M_{phys} = \frac{M(\beta)}{\epsilon(\beta)}
  = \frac{M(\beta)}{\sqrt{\sigma(\beta)/\sigma_{\rm phys}}}
  = \frac{M(\beta)}{\sqrt{\sigma(\beta)}} \sqrt{\sigma_{\rm phys}} .
\end{align}
The magnetic condensations of mass dimension two are translated as
\begin{align}
  \left< (\partial_\rho \bm{n}(x))^2 \right>_{phys} 
  = \frac{\left< (\Delta_\rho \bm{n}(x))^2 \right>(\beta)}{\epsilon^2(\beta)}
  = \frac{\left< (\Delta_\rho \bm{n}(x))^2 \right>(\beta)}{\sigma(\beta)} \sigma_{\rm phys} .
\end{align}

 Both magnetic condensations of mass dimension two increase monotonically as the lattice spacing $\epsilon$ decreases (or $\beta$ increases).


%

\subsection{The ratio}


\begin{figure}[htbp]
\begin{center}
\includegraphics[height=6.0cm]{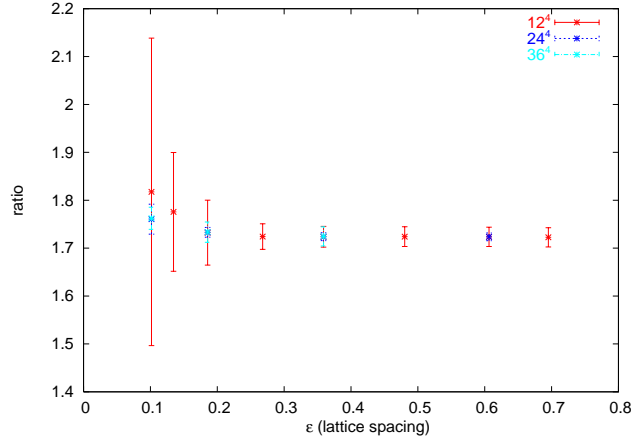}
\caption{\small The ratio $r:=\left<g^2 \mathbb{B}^2\right>/\left< g\| \mathbb{H} \| \right>$ of two condensates 
versus the lattice spacing $\epsilon$ on $12^4, 24^4, 36^4$ lattices. 
}
\label{fig:ratio}
\end{center}
\end{figure}



%

The precise ratio $r:=\left<g^2 \mathbb{B}^2\right>/\left< g\| \mathbb{H} \| \right>$ between two magnetic condensations is plotted in  Fig.~\ref{fig:ratio}.  
Although the respective condensation changes considerably with decreasing in the lattice spacing $\epsilon$ (or increasing $\beta$), the ratio converges to a value $r \sim 1.8$ in the continuum limit $\epsilon \downarrow 0$. 
The obtained value of the ratio $r \sim 1.8$ supports the recovery of stability of the Savvidy vacuum according to the argument \cite{Kondo04}.

The fact that the increase of two vacuum condensates and the constancy of the ratio with respect to $\beta$  suggests that the composite operators
$g^2 \mathbb{B}^2$ and $g\| \mathbb{H} \|$ besides the field $\bm{n}$ have non-zero anomalous dimensions which are nearly equal to each other. 
The anomalous dimension of the field $\bm{n}$ is obtained by calculating  the correlation function 
$\left< n^A(x) n^B(y) \right>$ in the short distance $|x-y|$ or high energy-momentum region, just as obtained in the non-linear sigma model in two dimensions which has the asymptotic freedom \cite{Polyakov75,PS95}.  
The numerical determination of the anomalous dimension of the composite operator is possible in principle.
However, this is still beyond the ability of our numerical calculations and to be reserved as a future problem.

\subsection{Lattice effective potential}

The probability distribution of the local operator $\Phi(x)$ is obtained by calculating  the expectation value 
\begin{align}
  \left< \delta(\varphi - \Phi(x)) \right> .
\end{align}
The effective potential is obtained from this distribution by taking the logarithm and changing the signature \cite{CPV97},
\begin{align}
  V_{eff}(\varphi) = - \ln \left< \delta(\varphi - \Phi(x)) \right> .
\end{align}
The effective constraint potential \cite{GL91} is defined for the averaged operator $\bar{\Phi}:=V^{-1} \sum_{x \in V} \Phi(x)$ over the four-volume $V$ by 
\begin{align}
  V_{eff}(\varphi) = - \ln \left< \delta \left(\varphi - V^{-1} \sum_{x \in V} \Phi(x) \right) \right> .
\end{align}
The value of the composite field, at which the potential has a minimum or the field distribution is maximum, is equal to the value of the vacuum condensate. 
This argument can be easily extended to a number of operators, $\Phi_1, \Phi_2, \cdots$ and the effective potential $V_{eff}(\varphi_1, \varphi_2, \cdots)$.


\begin{figure}[htbp]
\begin{center}
\begin{picture}(450,210)
\put(-40,-20){
 \includegraphics[height=8.5cm]{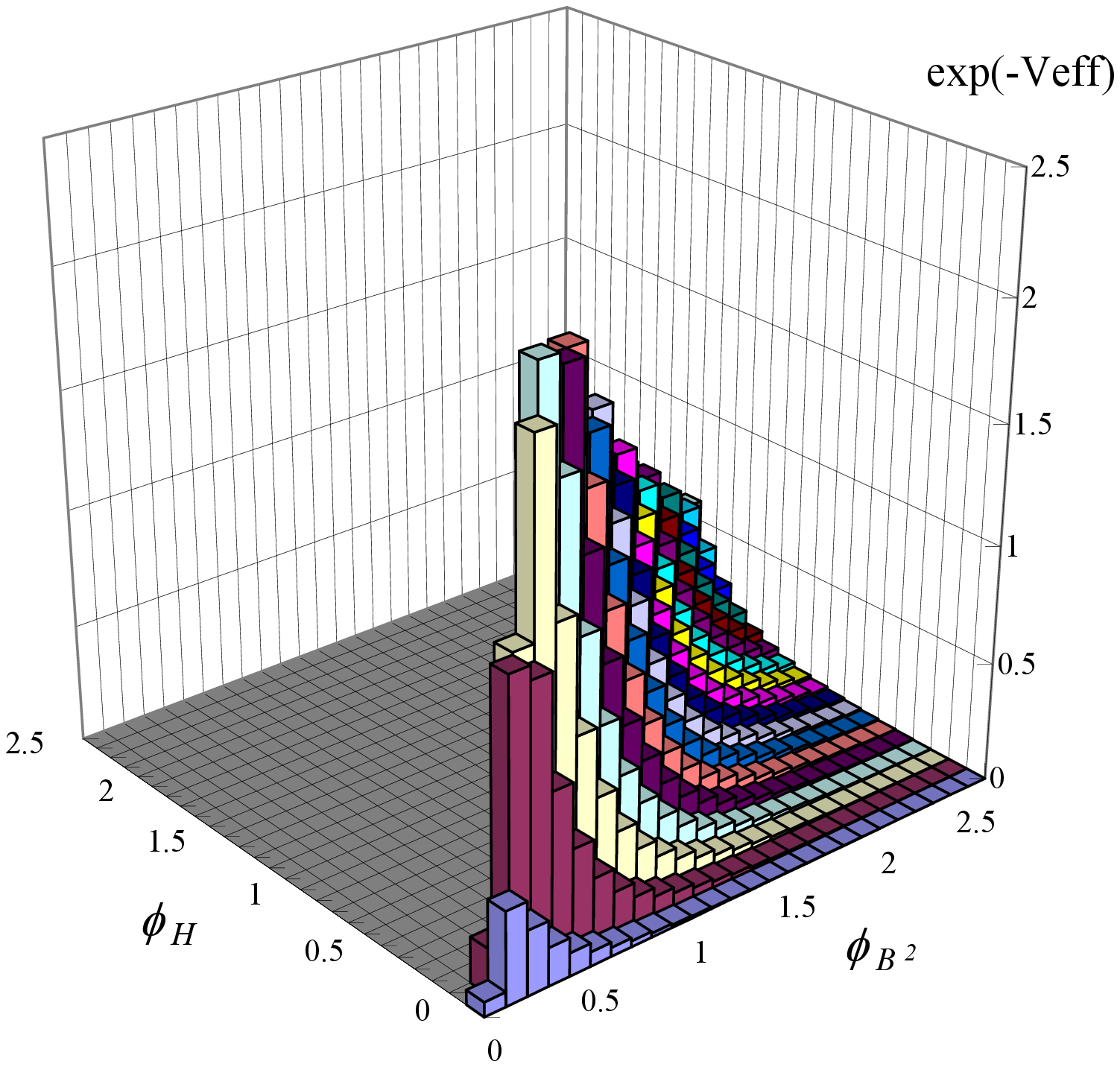}
}
\put(190,-20){
 \includegraphics[height=8.5cm]{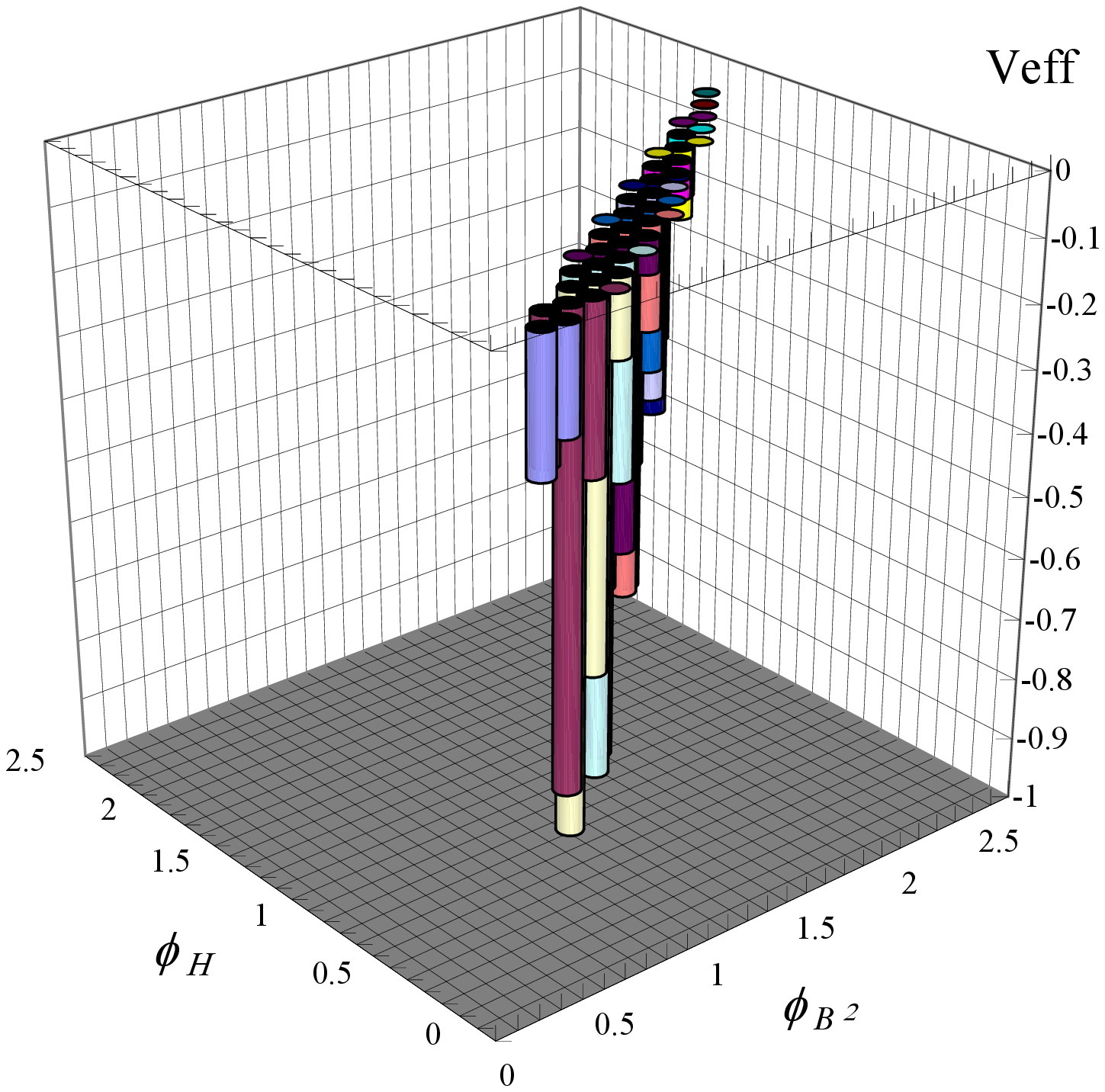}
}
\end{picture}
\caption{\small 
The local composite operators 
$\phi_{B^2}(x):=\mathbb{B}^2(x)$ and $\phi_{H}(x):=\|\mathbb{H}(x)\|$ 
obey
 (left panel) the probability distribution independent of $x$: 
 $\exp [-V_{eff}(\phi_{B^2}, \phi_{H})]$ , 
 and
 (right panel) the effective potential $V_{eff}(\phi_{B^2}, \phi_{H})$,
 at $\beta=2.3$ on $24^4$ lattice for 500 samples. 
}
\label{fig:local-eff}
\end{center}
\end{figure}



\begin{figure}[htbp]
\begin{center}
\begin{picture}(450,210)
\put(-40,-20){
 \includegraphics[height=8.5cm]{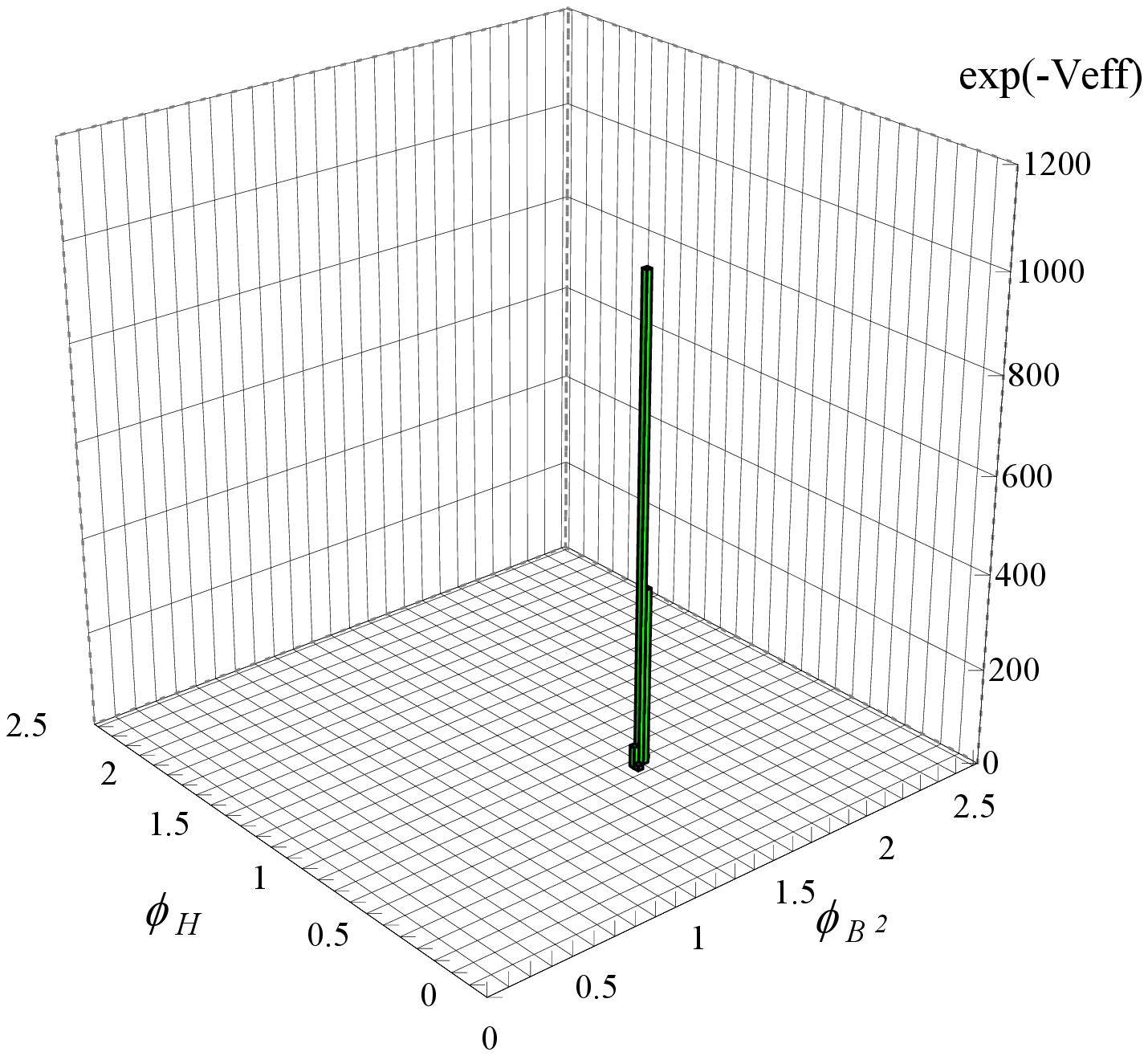}
}
\put(190,-20){
 \includegraphics[height=8.5cm]{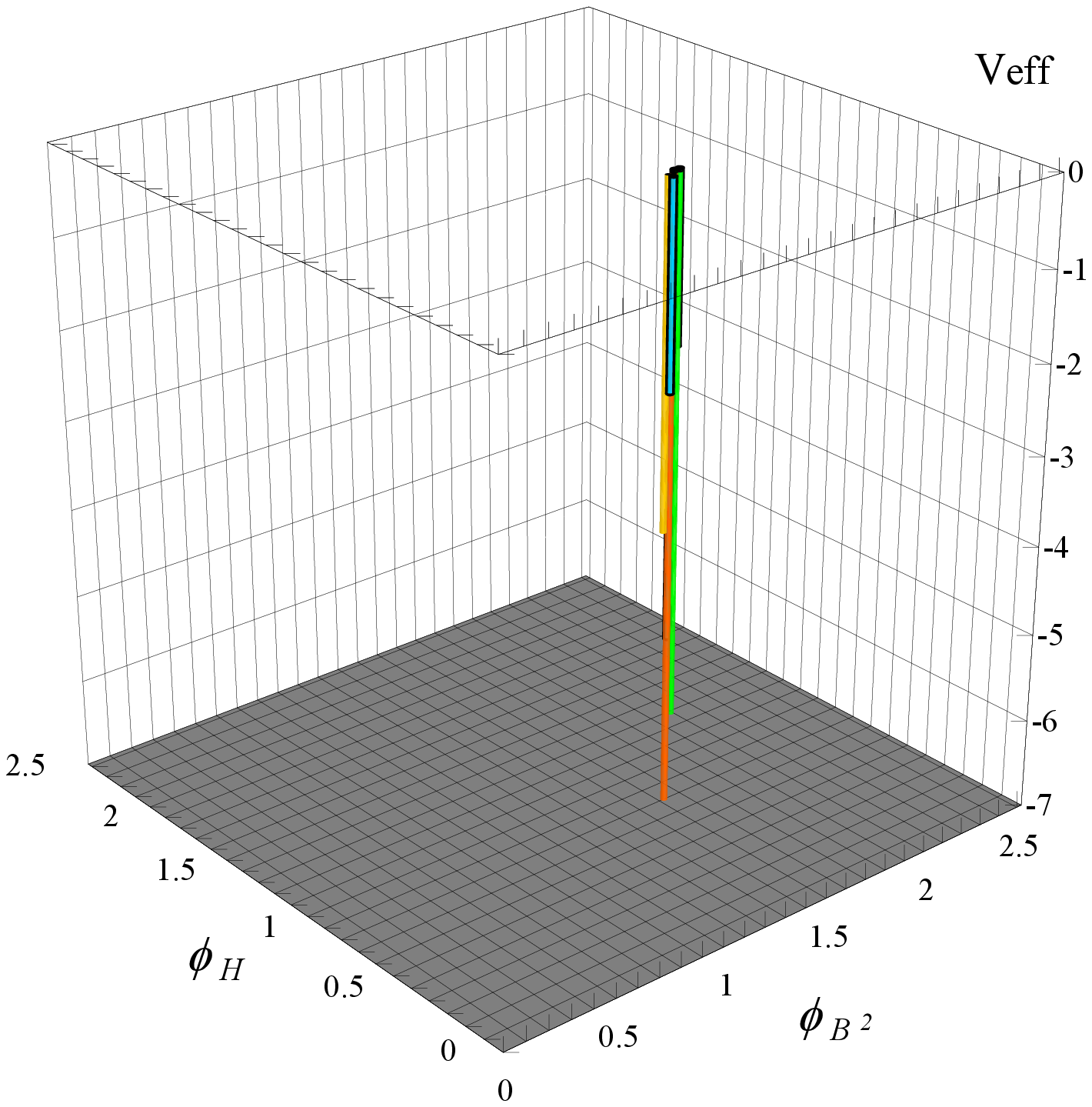}
}
\end{picture}
\caption{\small
The averaged composite operators 
$\phi_{B^2}:= V^{-1} \sum_{x \in V} \mathbb{B}^2(x)$
and 
$\phi_{H}:= V^{-1} \sum_{x \in V} \|\mathbb{H}(x)\|$
obey 
 (left panel) the probability distribution: 
 $\exp [-V_{eff}(\phi_{B^2}, \phi_{H})]$, 
 and 
 (right panel) the constraint effective potential $V_{eff}(\phi_{B^2}, \phi_{H})$,
 at $\beta=2.3$ on $24^4$ lattice for 500 samples. 
}
\label{fig:average-eff}
\end{center}
\end{figure}


In our case, we can define two effective potentials written in terms of the values of two  composite operators:  
\begin{align}
  V_{eff}(\phi_{B^2},\phi_{H}) 
  = - \ln \left< \delta(\phi_{B^2} - g^2\mathbb{B}^2(x)) \delta(\phi_{H} - g\|\mathbb{H}(x)\|) \right> ,
  \label{localpot}
\end{align}
and
\begin{align}
  V_{eff}(\phi_{B^2},\phi_{H})
   =& - \ln \left< \delta \left(\phi_{B^2} - V^{-1} \sum_{x \in V} g^2\mathbb{B}^2(x) \right)
    \delta \left(\phi_{H} - V^{-1} \sum_{x \in V} g\|\mathbb{H}(x)\| \right) \right> .
    \label{avepot}
\end{align}
See Fig.~\ref{fig:local-eff} and Fig.~\ref{fig:average-eff} for the effective potentials obtained in the LLG and SU(2)$_{global}$-invariant MAG.  
Our simulations  have shown that the local potential (\ref{localpot}) is independent of the point $x$ and hence the spacetime average of the local potential is plotted in Fig.~\ref{fig:local-eff}.

The numerical calculations show that 
the support of $V_{eff}(\phi_{B^2},\phi_{H})$ and the distribution are contained in the allowed region 
$g^2\mathbb{B}^2 > g \| \mathbb{H} \|$  
and that the minimum of $V_{eff}(\phi_{B^2},\phi_{H})$ and the maximum of the distribution are indeed shifted from zero in the allowed region. 
These results clearly indicate the simultaneous existence of two vacuum condensates, although two operators $\mathbb{B}^2$ and $\| \mathbb{H} \|$ are always greater than or equal to zero.%
\footnote{  
We can see that the distribution of $\| \mathbb{H} \|$ in Fig.~\ref{fig:local-eff} is consistent with the value of the standard deviation $\sigma$ calculated from date of Fig.~\ref{fig:HH} according to 
$\sigma:=\sqrt{\langle (\|g\mathbb{H}\|-\langle \|g\mathbb{H}\| \rangle)^2 \rangle}
= \sqrt{\langle  \|g\mathbb{H}\|^2 \rangle -\langle  \|g\mathbb{H}\| \rangle^2 }$,
 e.g., $\sigma \cong \sqrt{1.028-0.867^2} \cong 0.526$ at $\beta=2.3$. 
}
In the deconfinement phase, the minimum is expected to be at the zero value of the composite operators $\mathbb{B}^2$ and $\| \mathbb{H} \|$.

Thus the numerical results obtained in this paper confirm the  qualitative result obtained by analytical calculations to the one-loop level in the previous paper \cite{Kondo04}.

\subsection{Lorentz invariance on a lattice}


\begin{figure}[htbp]
\begin{center}
\includegraphics[height=5.0cm]{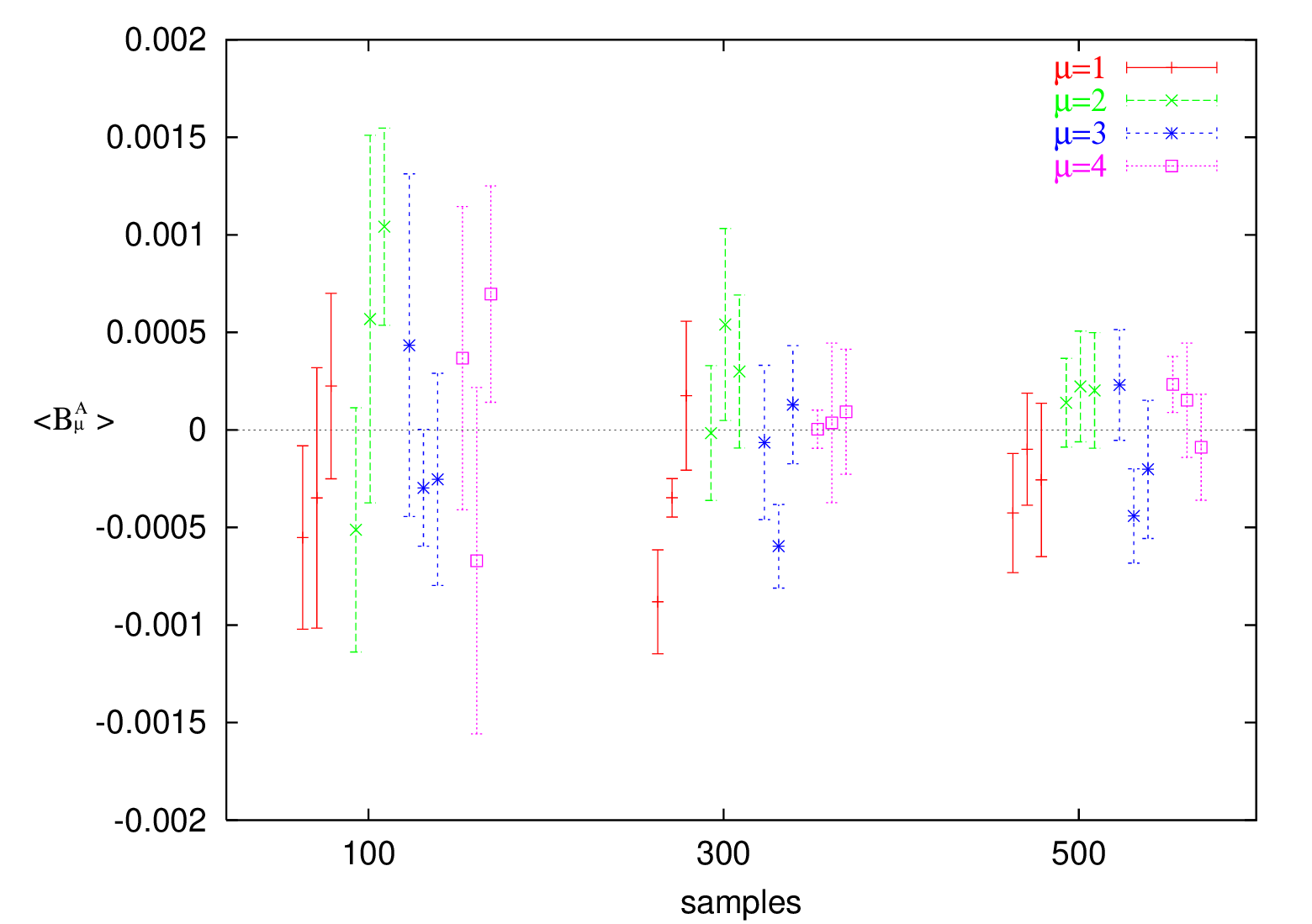}
\includegraphics[height=5.0cm]{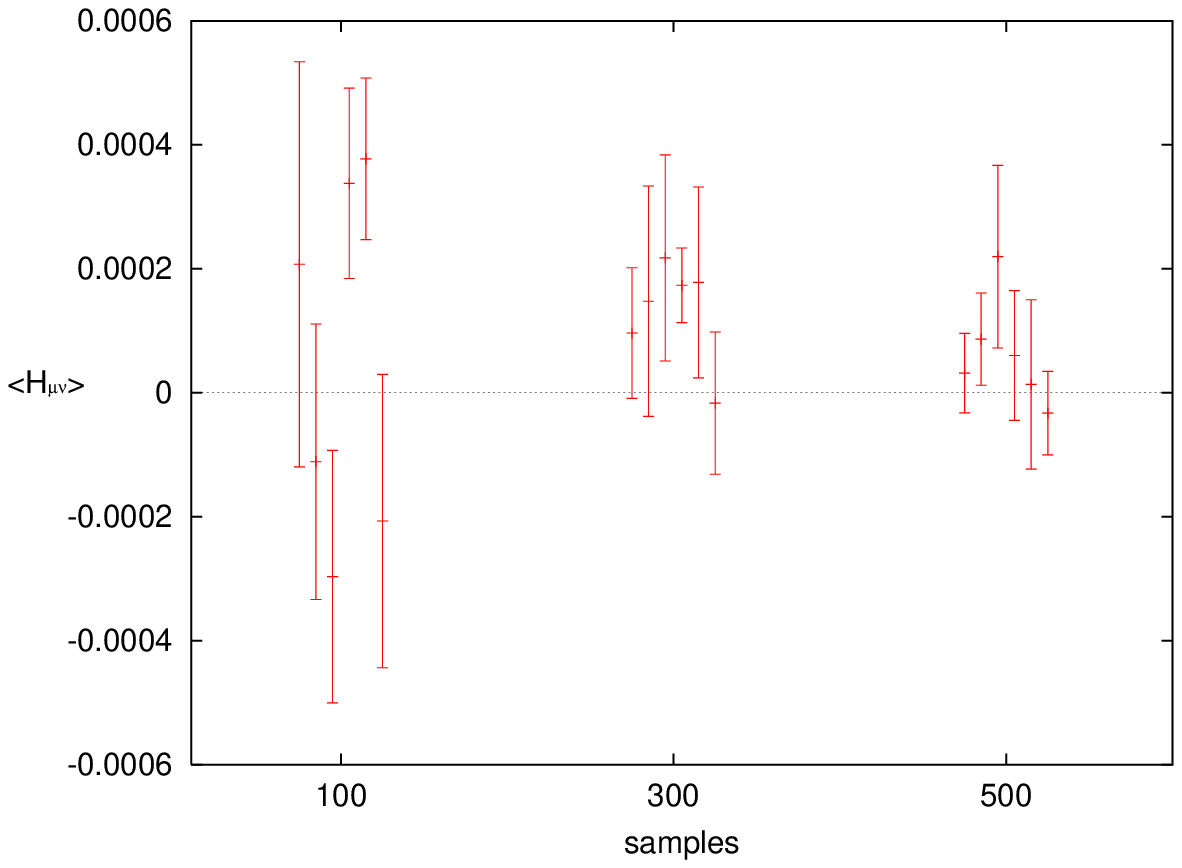}
\caption{\small
Check of Lorentz invariance. Plots of  
(left panel) $V^{-1}\sum_{x \in V} \left\langle B_{x,\mu}^{A}\right\rangle$ 
for $\mu=1,2,3,4$ and $A=1,2,3$, 
(right panel) $V^{-1}\sum_{x \in V}  \left\langle H_{x,\mu \nu}\right\rangle$ 
for $\mu,\nu=1,2,3,4$, on $24^{4}$ lattice at $\beta =2.3$ using 100, 300
and 500 samplings.
}
\label{fig:B}
\end{center}
\end{figure}


The magnetic condensations measured so far are defined in the Lorentz invariant way from the beginning. 
On the lattice, the Lorentz invariance (Euclidean rotational invariance) is inevitably broken due to a non-zero lattice spacing. 
However, the rotation invariance by angle $\pi/2$ exists even on the isotropic lattice.
To see this discrete rotation invariance, we have measured the vacuum expectation values of a component of the Lorentz vector $\left\langle B_{x,\mu }^{A}\right\rangle $ and  the Lorentz tensor $\left\langle H_{x,\mu \nu }\right\rangle =\left\langle \bm{n}_{x}\cdot \mathbb{H}_{x,\mu \nu }\right\rangle $ on a lattice, see  
 Fig.~\ref{fig:B}. 
What they vanish is a necessary condition for the full Lorentz invariance of the continuum theory $\left\langle B_{\mu}^A(x)\right\rangle =0$ and $\left\langle H_{\mu \nu }(x)\right\rangle =0$.
The vacuum expectation values 
$\left\langle B_{x,\mu }^{A}\right\rangle $ and $\left\langle H_{x,\mu \nu }\right\rangle$ on a lattice 
should be zero reflecting the discrete rotation invariance.
In fact, Fig.~\ref{fig:B} indicates that the respective component is extremely small compared to  the relevant vacuum condensates 
$\left< \mathbb{B}_{x,\mu}^2 \right>$ and $\left< \sqrt{\mathbb{H}_{x,\mu\nu}{}^2} \right>$ defined in the Lorentz invariant (and global gauge invariant) way. 
Moreover, it is observed that the absolute value with the error are decreasing monotonically, as the number of samplings is increasing, as expected. 

Thus we conclude that the magnetic condensations and the dynamical generation of color magnetic field presented in this paper do not mean the violation of the Lorentz invariance in the Yang-Mills theory.  This result is in sharp contrast with the original Savvidy and Copenhagen vacuum.

\subsection{Lattice Gribov copies}


\begin{figure}
\begin{center}
\includegraphics[height=5.5cm]{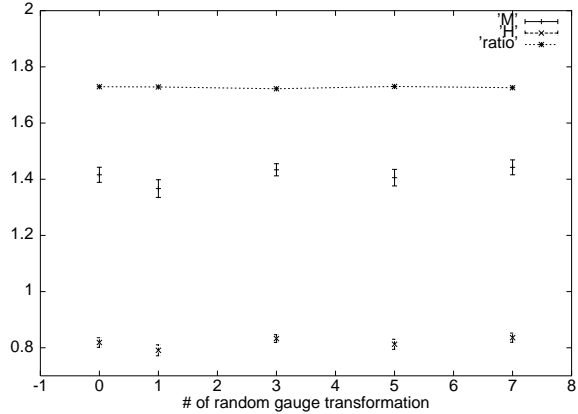}
\caption{\small 
The effect of lattice Gribov copies:  $r$, $\left<g^2 \mathbb{B}^2\right>$ and $\left< g\| \mathbb{H} \| \right>$ (from up to down) versus  the number of how many times the random gauge transformations are performed for the original configurations, at $\beta$=2.35 with the lattice size $L= 8^4$ for 
thermalization=3000, iteration=100, number of configurations=30.
}
\label{fig:gribovcopy}
\end{center}
\end{figure}



In our calculations of magnetic condensations, 
we have also estimated the effect of lattice Gribov copies due to the program of performing the gauge fixing on a lattice \cite{BBMPS96,IKPS03}.
We have used a standard iterative gauge fixing procedure for MAG and
LLG. In such a case, gauge fixing sweeps may be stuck for
some local minima of a gauge fixing functional.  Different local
minima give rise to different gauge transformations, but they can not
be distinguished from the viewpoint of the iterative gauge fixing
procedure. These are the lattice Gribov copies. 
To check the effect of copies to the magnetic condensations, 
we generate 30 of $SU(2)$ configurations $\{U_{x,\mu}\}$
on $8^4$ lattice at $\beta=2.35$.  Then, we generate 4 of gauge
equivalent configurations (i.e., copies) via a random gauge
transformation before performing the LLG.  
Using these gauge copies, we estimated the novel type of 
vacuum condensation $\left<\mathbb{B}_{\mu}^2\right>$, the magnetic condensation 
$\left<||\mathbb{H}||\right>$ and the index for the stability restoration 
$r=\left<g^2\mathbb{B}_{\mu}^2\right>/\left<g||\mathbb{H}||\right>$.
Fig.~\ref{fig:gribovcopy} 
shows the ratio $r$, $\left<g^2\mathbb{B}_{\mu}^2\right>$,and $\left<g||\mathbb{H}||\right>$. 
The horizontal axis show the number of times of a random 
gauge transformation.

This result shows that the ratio is stable, although the respective 
condensation is a little affected by Gribov copies.
Therefore, qualitative analyses given in this 
section will not be affected by Gribov copies.

\subsection{Anatomy of dimension two condensates}

We define the spacetime average of the vacuum expectation value  for the local operator $\mathcal{O}(x)$ by
\begin{align}
  \langle\langle \mathcal{O} \rangle\rangle := \frac{1}{\Omega} \int_{\Omega} d^4x \left< \mathcal{O}(x) \right> . 
\end{align}

Using the CFN variable, 
$
  \mathscr{A}_\mu(x) 
  = \mathbb{V}_\mu(x)   + \mathbb{X}_\mu(x)  
  = \mathbb{C}_\mu(x) + \mathbb{B}_\mu(x) + \mathbb{X}_\mu(x) , 
$
the squared potential $\langle\langle \mathscr{A}_\mu^2 \rangle\rangle$ is decomposed as
\begin{align}
  \langle\langle \mathscr{A}_\mu^2 \rangle\rangle
  &=  \langle\langle \mathbb{V}_\mu^2 \rangle\rangle
  + 2 \langle\langle \mathbb{V} \cdot \mathbb{X} \rangle\rangle 
  + \langle\langle \mathbb{X}_\mu^2 \rangle\rangle 
  \nonumber\\
  &= \langle\langle \mathbb{C}_\mu^2 \rangle\rangle 
  + \langle\langle \mathbb{B}_\mu^2 \rangle\rangle
  + 2 \langle\langle \mathbb{B} \cdot \mathbb{X} \rangle\rangle 
  + \langle\langle \mathbb{X}_\mu^2 \rangle\rangle
  .
\end{align}
Following Zakharov et al. \cite{GSZ01}, the minimum of the squared potential $\langle\langle \mathscr{A}_\mu^2 \rangle\rangle$ with respect to the gauge transformation is gauge invariant.
It should be remarked that the operator $\mathbb{X}_\mu^2$ is  gauge invariant under the  SU(2) local gauge transformation II, see \cite{KMS05}.  Therefore, the difference 
$\langle\langle \mathscr{A}_\mu^2 \rangle\rangle - \langle\langle \mathbb{X}_\mu^2 \rangle\rangle$
could have a gauge invariant meaning. 
\footnote{ 
Recently,  there appeared many papers \cite{GSZ01,Kondo01,Boucaudetal00,ABB04,Slavnov04,Slavnov05,Kondo05} discussing the gauge invariance for the spacetime average of mass dimension two condensate
$\langle \mathscr{A}_\mu^2 \rangle$    i.e., 
$\langle\langle \mathscr{A}_\mu^2 \rangle\rangle$.
Among them, Slavnov\cite{Slavnov04,Slavnov05,Kondo05} claims a stronger statement that $\langle\langle \mathscr{A}_\mu^2 \rangle\rangle$ can be gauge invariant and can have the same value independent of the gauge fixing adopted. 
}


\begin{figure}[htbp]
\begin{center}
\includegraphics[height=5.0cm]{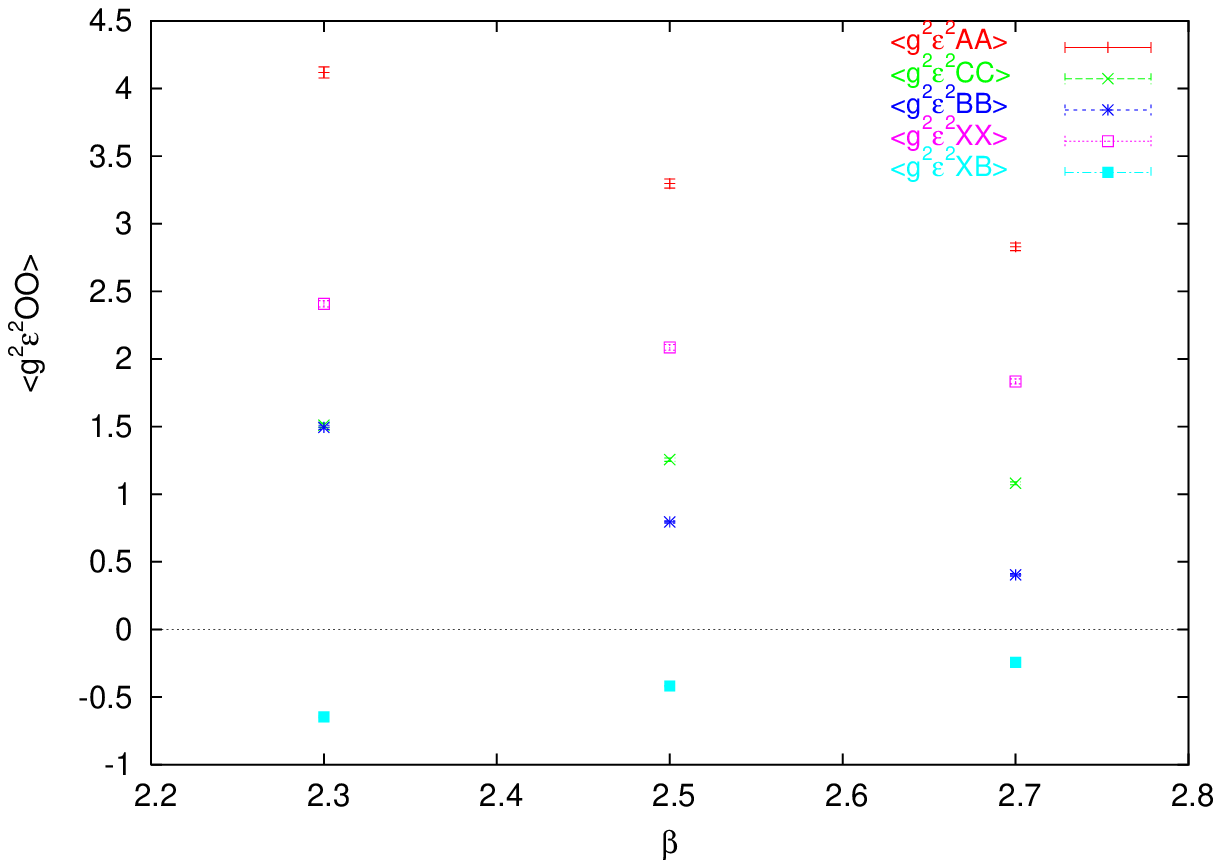}
\includegraphics[height=5.0cm]{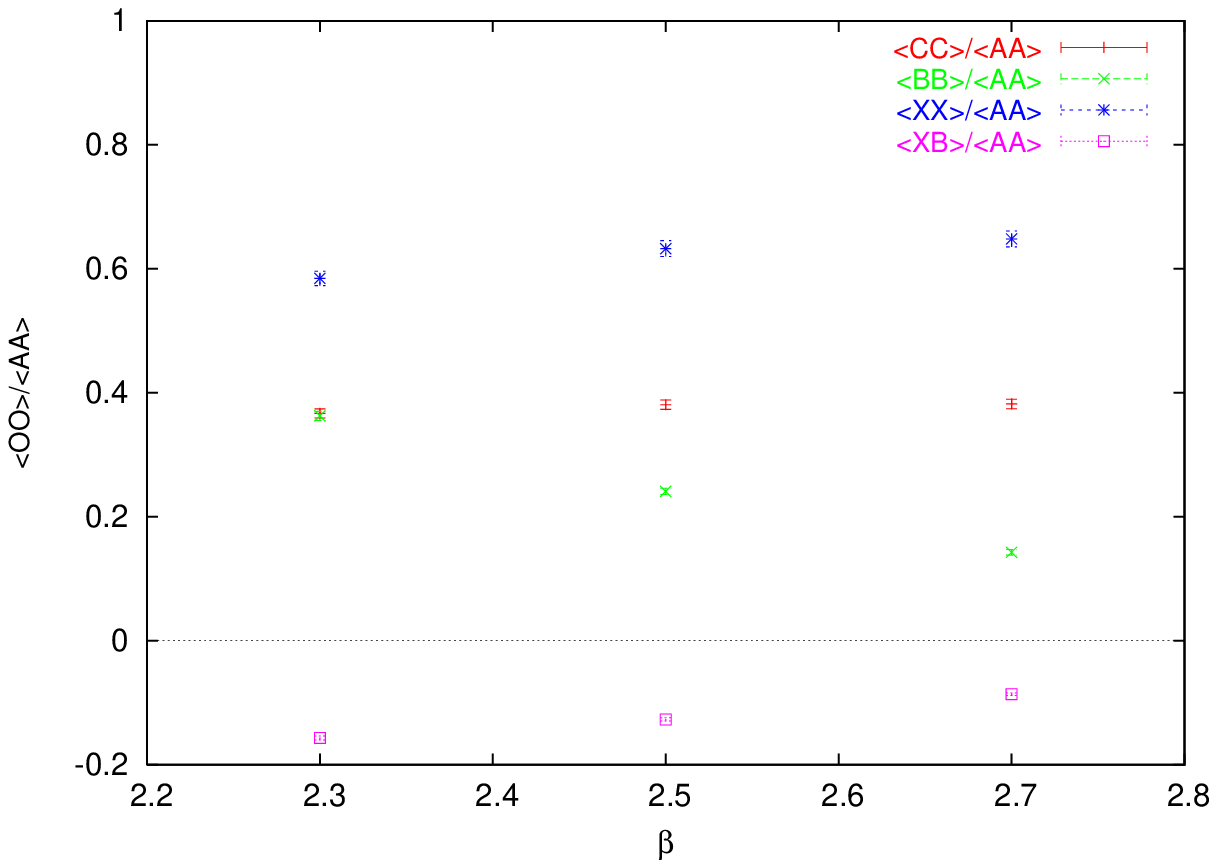}
\caption{\small 
Anatomy of dimension two condensates. From top to down, positive 
$\langle\langle \epsilon^2 g^2 \mathscr{A}_\mu^2 \rangle\rangle$, 
$\langle\langle \epsilon^2 g^2\mathbb{X}_\mu^2 \rangle\rangle$, 
$\langle\langle \epsilon^2 g^2\mathbb{C}_\mu^2 \rangle\rangle$, 
$\langle\langle \epsilon^2 g^2\mathbb{B}_\mu^2 \rangle\rangle$, 
and negative $\langle\langle \epsilon^2 g^2\mathbb{X}_\mu \cdot \mathbb{B}_\mu \rangle\rangle$.
(Left panel) the bare values measured on a lattice, (Right panel) the ratio of the respective dimension two condensate $\langle\langle \mathscr{O}^2 \rangle\rangle$ to the original dimension two condensate $\langle\langle \mathscr{A}_\mu^2 \rangle\rangle$. 
The presented values are raw data, i.e., bare values without any renormalizations. 
In the simulations, the gauge potential $\mathscr{A}_\mu(x)$ is extracted from the link variable $U_{x,\mu}$ using the linear definition, i.e, 
$\epsilon g\mathscr{A}_\mu(x) = \frac{i}{2}(U_{x,\mu}-U_{x,\mu}^\dagger)$.
The simulations are performed on the $24^4$ lattice under the same conditions as in the other simulations. 
}
\label{fig:sym-cfn-ym}
\end{center}
\end{figure}


Note that the cross term $\langle\langle \mathbb{V} \cdot \mathbb{X} \rangle\rangle=\langle\langle \mathbb{B} \cdot \mathbb{X} \rangle\rangle$ is eliminated by a special choice of the gauge transformation II,  
$\bm{\omega}' = \omega_{\parallel} \bm{n}$  (residual U(1) invariance) even after the nMA gauge is imposed, and hence 
$
  \langle\langle \mathscr{A}_\mu^2 \rangle\rangle
  \rightarrow 
  \langle\langle c_\mu^2 \rangle\rangle 
  + \langle\langle \mathbb{B}_\mu^2 \rangle\rangle
  + \langle\langle \mathbb{X}_\mu^2 \rangle\rangle 
=  \langle\langle \mathbb{V}_\mu^2 \rangle\rangle
  + \langle\langle \mathbb{X}_\mu^2 \rangle\rangle . 
$
However, it does no longer hold in general after the complete gauge fixing.

It is pointed out in \cite{Kondo04} that the off-diagonal gluon condensation of mass dimension two $\langle\langle \mathbb{X}_\mu^2 \rangle\rangle \not= 0$ leads to the Skyrme--Faddeev model as a low-energy effective theory of Yang-Mills theory.
The existence was also assumed as a key ingredient in the studies \cite{FN02,NW05}.

In view of these, we have measured various dimension two condensates constructed from the CFN variables,  including 
the off-diagonal gluon condensation $\langle\langle \mathbb{X}_\mu^2 \rangle\rangle \not= 0$, for the first time, see Fig.~\ref{fig:sym-cfn-ym}. 
Here we have performed the CFN decomposition on a lattice according to (\ref{LCFN}).  
The numerical simulations show that the cross term $\langle\langle \mathbb{V} \cdot \mathbb{X} \rangle\rangle=\langle\langle \mathbb{B} \cdot \mathbb{X} \rangle\rangle$ does not vanish and becomes negative, 
$\langle\langle \mathbb{B} \cdot \mathbb{X} \rangle\rangle<0$,
after the complete gauge fixing, i.e., the Landau gauge fixing in addition to the nMAG. 
In other words, $\mathbb{X}_\mu$ is not in the mass eigenstate $\tilde{\mathbb{X}}_\mu$. 
However, the  condensate $\langle\langle \tilde{\mathbb{X}}_\mu^2 \rangle\rangle = \langle\langle \mathbb{X}_\mu^2 \rangle\rangle \not= 0$ has the same value.

\section{Conclusion and Discussion}
\setcounter{equation}{0}

We have implemented the Cho-Faddeev-Niemi decomposition in the SU(2) Yang-Mills theory on a lattice.
Performing the Monte Carlo simulation on a lattice based on this framework, we have obtained a first numerical evidence for the existence of a novel magnetic condensation $\left<\mathbb{B}_\mu^2 \right>$ in addition to another magnetic condensation $\left< \|\mathbb{H}\| \right>$ corresponding to the Savvidy-like magnetic field. 
We have confirmed the existence of the vacuum condensations by calculating the effective potential on a lattice and obtained the stable value for the ratio $r \sim 1.8$ in favor of stability restoration of the Savvidy vacuum according to the previous paper \cite{Kondo04}. 
Moreover, it has been checked that the magnetic condensations in question do  not break the Lorentz invariance.

In the previous paper \cite{Kondo04}, we have argued that the stability of the Savvidy vacuum is restored due to the dynamical mass generation of off-diagonal gluons  caused by a novel type of magnetic condensation (with mass dimension two) coming from magnetic monopole degrees of freedom.
The off-diagonal gluons acquire the dynamical mass, 
$M_X^2=g^2 \left<\mathbb{B}_\mu^2 \right>$
due to the existence of  
 a novel magnetic condensation $\left<\mathbb{B}_\mu^2 \right>$  and removes the tachyon mode of the off-diagonal gluon to cure the Nielsen--Olesen instability of the Savvidy vacuum, while the diagonal gluon remains massless. 
To really confirm this claim, 
we must check whether the off-diagonal gluon mass $M_X$ determined by measuring  the decay rate of the correlation function 
$\left< X^A_\mu(x) X^B_\nu(y) \right>$ agrees with the magnetic condensation
$g^2 \langle \mathbb{B}_\mu^2 \rangle$. 
In order to know the absolute value of the condensate 
$\left<\mathbb{B}_\mu^2 \right>$, we need to know  more detailed behaviors of the propagator, e.g., the anomalous dimension of the field $\bm{n}$. 
Analytical attempt of calculating the anomalous dimension is now in progress within the continuum formulation.  

The other vacuum condensation 
$\langle \mathbb{X}_\mu^2 \rangle$ 
is also important.
In fact, the {\it off-diagonal gluon condensation of mass dimension 2} proposed in the MAG \cite{Kondo01},   
$
\langle  \mathbb{X}_\rho \cdot \mathbb{X}_\rho  \rangle
= 2\Lambda^2 \not= 0 ,
$
in the present framework
yields the {\it mass term} for the field $\mathbb{B}_\mu$ or the kinetic term for $\bm{n}$ through the interaction term 
$\frac{1}{2} \mathbb{B}_\mu \cdot \mathbb{B}_\mu \mathbb{X}_\rho \cdot \mathbb{X}_\rho$:
\begin{align}
  g^2 \Lambda^2  \mathbb{B}_\mu \cdot \mathbb{B}_\mu 
=    \Lambda^2  (\partial_\mu \bm{n})^2  . 
\end{align}
Therefore, the off-diagonal gluon condensation yields the Skyrme-Faddeev  model \cite{FN97}, which has been proposed as a low-energy effective theory of Yang-Mills theory and is supposed to describe the glueball by the knot soliton solution.

A  way to obtain a gauge invariant characterization of dual superconductivity in QCD is to calculate the Wilson loop average.
This is in principle possible in the same framework using the CFN decomposition based on a version of the non-Abelian Stokes theorem \cite{DP89,KondoIV}.  

It is also important to clarify the relationship between magnetic condensation discussed in this paper and magnetic monopole condensation as a source of dual superconductor, in order to confirm the magnetic monopole dominance.  
The issues are to be reported in subsequent papers. 

\vskip 0.5cm
{\bf Remarks on the gauge invariance}

The numerical simulations performed in this paper are based on a new interpretation of the CFN decomposition proposed in the paper \cite{KMS05}. As shown in \cite{KMS05}, in order to determine the configurations of the $\bm{n}$ field, the complete gauge fixing is needed, and we have adopted the nMAG and the Landau gauge in this simulation. In this sense, the obtained configurations of $\bm{n}$ field depend on the gauge adopted. 

However, the following points should be remarked. 
\begin{enumerate}
\item
We can choose an arbitrary gauge-fixing condition other than the Landau to fix the SU(2) local gauge symmetry II which remains after the nMAG, as clarified in \cite{KMS05}. 

\item
The $\bm{n}$ field is not a directly measurable physical quantity, since it is a vector indicating the color direction at each spacetime point and is not a color singlet object.  Therefore, even if the $\bm{n}$ field is subject to changes by taking different gauge fixing conditions, it does not cause the observable phenomena.  Hence, this does not lead to any difficulty. 
\end{enumerate}

Therefore, the problem is whether or not the resulting changes of the $\bm{n}$ field influence the magnetic condensations in question.  As already mentioned in ref.\cite{Kondo04}, we know that the operators $\mathbb{B}_\mu^2$ and $\sqrt{\mathbb{H}_{\mu\nu}^{2}}$ whose expectation values are measured in this paper are not invariant under the full SU(2) local gauge transformation II, although they are invariant under the global gauge transformation II (color rotation).  (After the nMAG, the theory has the SU(2) local gauge symmetry II, see \cite{KMS05}).  
In the operator level, the full SU(2) gauge invariant combinations are obtained by including the electric components as 
$\sqrt{(\mathbb{E}_{\mu\nu}+{\mathbb{H}_{\mu\nu}})^2}$ 
or
$\mathbb{V}_\mu^2=(\mathbb{B}_\mu+\mathbb{C}_\mu)^2=\mathbb{B}_\mu^2+\mathbb{C}_\mu^2$.
Therefore, from the viewpoint of gauge invariance in the operator level, we should have measured the quantities
$\left< \sqrt{(\mathbb{E}_{\mu\nu}+{\mathbb{H}_{\mu\nu}})^2} \right>$ or $\left< \mathbb{V}_\mu^2 \right>$.  Nevertheless, we have measured only $\left< \mathbb{B}_\mu^2 \right>$ or $\left< \sqrt{\mathbb{H}_{\mu\nu}^{2}} \right>$ in this paper and  avoided including the electric components. The reasons are as follows. 

\begin{enumerate}
\item
The dimension two composite operators $\mathbb{B}_\mu^2$, $\sqrt{\mathbb{H}_{\mu\nu}^2}$, and the dimension four operators  $\mathbb{H}_{\mu\nu}^2$ are gauge invariant in the operator level under the local U(1)$_{\rm II}$ gauge transformation, see Appendix \ref{Appendix:gt}. 
This is the same setting as the original approach of Nielsen--Olesen \cite{NO78}. 
Moreover, they are also color singlets, i.e., invariant under the global SU(2) gauge transformation II (color rotation).

\item
In this paper we are interested in the magnetic contributions coming from the topological degrees of freedom expressed through the $\bm{n}$ field.  

\item
According to the conventional wisdom, the pure color electric field makes the vacuum unstable due to gluon-antigluon pair annihilations in gluodynamics, just as the pure electric field makes the QED vacuum unstable due to electron-positron pair creations.  Therefore, the inclusion of the electric components $\mathbb{C}_\mu$ and $\mathbb{E}_{\mu\nu}$ under our identification could be other sources for the instability of the vacuum. This leads to difficulties in demonstrating our claim that the magnetic condensations stabilize the vacuum by eliminating the tachyon mode. 

\item
Our simulations are inspired by the analytical calculation \cite{Kondo04} to the one-loop order. In the pure magnetic case, the effective potential is obtained in the closed form. When the electric field and the magnetic field are simultaneously included, however, no one has succeeded to obtain the closed form for the effective potential and the known expression is a cumbersome infinite series.

\item
After taking spacetime average and the vacuum expectation value or Yang-Mills average, the gauge variant operator in the strong sense could become gauge invariant in the week sense as discussed in  \cite{Slavnov04,Slavnov05,Kondo05}. This leaves a possibility of SU(2) gauge invariance for $\left< \mathbb{B}_\mu^2 \right>$ and $\left< \sqrt{\mathbb{H}_{\mu\nu}^{2}} \right>$.

\item
Incidentally, the same type of a mathematical identity is applied to yield the lower bound on $\mathbb{V}_\mu^2$. But    the identity 
$
  (\mathbb{V}_\mu \cdot \mathbb{V}_\mu)^2 - (\mathbb{V}_\mu \times \mathbb{V}_\nu) \cdot (\mathbb{V}_\mu \times \mathbb{V}_\nu) = (\mathbb{V}_\mu \cdot \mathbb{V}_\nu) (\mathbb{V}_\mu \cdot \mathbb{V}_\nu) \ge 0, 
$
yields  not so beautiful and not so useful result:
\begin{align}
  g^2 \mathbb{V}_\mu^2 =g^2\mathbb{B}_\mu^2+g^2c_\mu^2 \ge g\sqrt{\mathbb{H}_{\mu\nu}^2+2[c_\mu^2 (\partial_\nu \bm{n})^2-c_\mu c_\nu \partial_\mu \bm{n} \cdot \partial_\nu \bm{n} ]} .
\end{align}


\end{enumerate}
Indeed, the effect of the electric field is important.  We have succeeded to separate all the CFN variables $\bm{n}$, $c_\mu$ and $\mathbb{X}_\mu$.  Therefore, we can now calculate the relevant quantities and estimate the desired electric contributions.  
Some of the results are presented in section 4.8 and  at the workshop\cite{Kondo-ringberg2005}.
We plan to perform the detailed investigation as the next work. 
However, we wish to avoid to present the details. Because, the inclusion of such materials makes the paper longer and the presentation could become rather incomplete. 
Without presenting the detailed numerical calculations, we can show that the other electric contributions increase the value of off-diagonal gluon mass and they are larger than the magnetic contributions, at least in the bare values before performing the renormalization by subtracting the perturbative ultraviolet divergent part.  Therefore, the inclusion of the electric part does not change the main claim of this paper: the tachyon mode is eliminated as a consequence of shifting the spectrum of the off-diagonal gluons upward due to the novel type of magnetic condensation (it is sufficiently achieved without the positive electric contribution).

\section*{Acknowledgments}
The numerical simulations have been done on a supercomputer (NEC SX-5) at  Research Center for Nuclear Physics (RCNP), Osaka University.
This work is also supported in part by
the Large Scale Simulation Program of High Energy Accelerator Research
Organization (KEK).
K.-I. K. is financially supported by 
Grant-in-Aid for Scientific Research (C)14540243 from Japan Society for the Promotion of Science (JSPS), 
and in part by Grant-in-Aid for Scientific Research on Priority Areas (B)13135203 from
the Ministry of Education, Culture, Sports, Science and Technology (MEXT).

\appendix
\baselineskip 14pt
\section{Gauge invariance and fixing in the CFN variable}
\setcounter{equation}{0}

\subsection{Gauge symmetry}

For the CFN decomposition,
\begin{equation}
\mathscr{A}_\mu(x)
 =c_\mu(x){\bm n}(x)
  +g^{-1}\partial_\mu{\bm n}(x)\times{\bm n}(x)
  +\mathbb X_\mu(x)
\end{equation}
the restricted potential $c_\mu$ and gauge covariant potential $\mathbb{X}_\mu$ are specified by $\bm{n}$ and $\mathscr{A}_\mu$:
\begin{align}
c_\mu(x)
 &={\bm n}(x)\cdot\mathscr{A}_\mu(x), 
\label{def:c}
\\
\mathbb X_\mu(x)
 &=g^{-1}{\bm n}(x)\times D_\mu[\mathscr{A}]{\bm n}(x) .
\label{def:X}
\end{align}
The second equation is obtained by making use of the fact that
\begin{align}
  D_\mu[\mathbb{V}] \bm{n} := \partial_\mu \bm{n} + g\mathbb{V}_\mu \times \bm{n} =  \partial_\mu \bm{n} + g\mathbb{B}_\mu \times \bm{n}
:= D_\mu[\mathbb{B}] \bm{n} \equiv 0 , 
\end{align}
which yields 
\begin{align}
  D_\mu[\mathscr{A}] \bm{n} = \partial_\mu \bm{n} +  g\mathscr{A}_\mu \times \bm{n} = \partial_\mu \bm{n} + g\mathbb{V}_\mu \times \bm{n} + g\mathbb{X}_\mu \times \bm{n} 
= g\mathbb{X}_\mu \times \bm{n} .
\label{XnA}
\end{align}
Therefore, the gauge transformations $\delta c_\mu$, $\delta\mathbb X_\mu$ are uniquely determined, once  the transformations 
$\delta{\bm n}$ and $\delta\mathscr{A}_\mu$ are specified. 
\begin{itemize}
\item
The fact ${\bm n}(x)^2=1$ urges us to consider the local rotation by an angle 
${\bm\theta}(x)$:
\begin{equation}
\delta{\bm n}(x)
  =g{\bm n}(x) \times {\bm\theta}(x) 
  =g{\bm n}(x) \times {\bm\theta}_\perp(x) ,
\end{equation}
where  
${\bm\theta}_\perp(x)$ are the perpendicular components of ${\bm\theta}(x)$ with two independent components ($\bm n\cdot{\bm\theta}=0$).
For the parallel component ${\bm\theta}_\parallel(x)=\theta_\parallel(x)\bm n(x)$, the vector field ${\bm n}(x)$ is invariant. Therefore, it is a redundant symmetry, which we call  U(1)$_{\theta}$ symmetry, of the Yang-Mills theory written in terms of CFN variables, since $c_\mu(x)$ and $\mathbb{X}_\mu(x)$ are also unchanged for a given $\mathscr{A}_\mu(x)$. 
This symmetry is the local SU(2)/U(1) symmetry and denoted by $[SU(2)/U(1)]_{local}^{\theta}$.

\item
The invariance of the Lagrangian is guaranteed by the usual gauge transformation:
\begin{equation}
\delta\mathscr{A}_\mu(x)
  =D_\mu[\mathscr{A}]{\bm\omega}(x) .
\end{equation}
\end{itemize}
This symmetry is the local SU(2) gauge symmetry and denoted by $SU(2)_{local}^{\omega}$. 

Note that 
${\bm\omega}(x)$ and ${\bm\theta}(x)$ are independent, since 
the original Yang-Mills Lagrangian is invariant irrespective of the choice of ${\bm\theta}(x)$.

For later convenience, we denote the above transformations by $\delta_\theta$ and $\delta_\omega$:
\begin{enumerate}
\item[(1)]
${\bm\theta}\ne{\bm0}$ ($\bm{n}\cdot{\bm\theta}=0$),
${\bm\omega}={\bm0}$:
\begin{equation}
\delta_\theta{\bm n}(x)
 =g{\bm n}(x)\times{\bm\theta}(x),
\quad
\delta_\theta\mathscr{A}_\mu(x)={\bm0}.
\label{gtn}
\end{equation}

\item[(2)]
${\bm\theta}={\bm0}$, ${\bm\omega}\ne{\bm0}$:
\begin{equation}
\delta_\omega{\bm n}(x)={\bm0},
\quad
\delta_\omega\mathscr{A}_\mu(x)
  =D_\mu[\mathscr{A}]{\bm\omega}(x).
\label{gtA}
\end{equation}

\end{enumerate}
Then the general gauge transformation of  the CFN variables is obtained by combining 
 $\delta_\theta$ and $\delta_\omega$.

\subsection{Local gauge transformations I and II}\label{Appendix:gt}

In the papers \cite{Cho03,Kondo04}, two local gauge transformations are introduced 
by decomposing the original gauge transformation, 
$
  \delta_\omega \mathscr{A}_\mu(x)  =  D_\mu[\mathscr{A}] \bm{\omega}(x) .
$
\footnote{
The gauge transformation I was called the passive or quantum gauge transformation, while II was called the active or background gauge transformation.  However, this classification is not necessarily independent, leading to sometimes confusing and misleading results.

}

\underline{Local gauge transformation I}:  
\begin{subequations}
\begin{align}
  \delta_\omega \bm{n}  =& 0  ,
\\
 \delta_\omega c_\mu =& \bm{n} \cdot  D_\mu[\mathscr{A}] \bm{\omega} ,
\\
  \delta_\omega \mathbb{X}_\mu =&     D_\mu[\mathscr{A}] \bm{\omega} - \bm{n}( \bm{n} \cdot  D_\mu[\mathscr{A}] \bm{\omega}) ,
\\ 
  \Longrightarrow  & \delta_\omega \mathbb{B}_\mu  =   0 ,
\quad
  \delta_\omega \mathbb{V}_\mu 
=   \bm{n}( \bm{n} \cdot  D_\mu[\mathscr{A}] \bm{\omega})  . 
\end{align}
\end{subequations}

\underline{Local gauge transformation II}:  
\begin{subequations}
\begin{align}
  \delta_{\omega}' \bm{n}  =& g \bm{n} \times \bm{\omega'}  ,
\\
 \delta_{\omega}' c_\mu =&    \bm{n} \cdot \partial_\mu \bm{\omega'}   ,
\\
  \delta_{\omega}' \mathbb{X}_\mu =&  g \mathbb{X}_\mu \times \bm{\omega'} ,
\\
\Longrightarrow  & \delta_{\omega}' \mathbb{B}_\mu  
=   D_\mu[\mathbb{B}] \bm{\omega'} - (\bm{n} \cdot \partial_\mu \bm{\omega'}) \bm{n}  ,
\quad
\delta_{\omega}' \mathbb{V}_\mu =   D_\mu[\mathbb{V}] \bm{\omega'}    . 
\end{align}
\end{subequations}

The gauge transformation for the field strength can be obtained in the similar way as follows. 

\noindent
\underline{Local gauge transformation I}:  
\begin{align}
  \delta_\omega \mathbb{E}_{\mu\nu} 
=& \bm{n} \delta_\omega {E}_{\mu\nu} 
= \bm{n} \{ \partial_\mu (\bm{n} \cdot D_\nu[\mathscr{A}] \bm{\omega}) - \partial_\nu (\bm{n} \cdot D_\mu[\mathscr{A}] \bm{\omega}) \} , 
\\
  \delta_\omega \mathbb{H}_{\mu\nu} 
=&  \bm{n} \delta_\omega {H}_{\mu\nu}  = 0 . 
\end{align}
\noindent
\underline{Local gauge transformation II}:  
\begin{align}
  \delta_\omega' \mathbb{E}_{\mu\nu} 
=& g \mathbb{E}_{\mu\nu} \times \omega' + \bm{n} \{ \partial_\mu (\bm{n} \cdot \partial_\nu \bm{\omega}') - \partial_\nu (\bm{n} \cdot \partial_\mu \bm{\omega}') \} , 
\\
  \delta_\omega' \mathbb{H}_{\mu\nu} =& g \mathbb{H}_{\mu\nu} \times \omega' - \bm{n} \{ \partial_\mu (\bm{n} \cdot \partial_\nu \bm{\omega}') - \partial_\nu (\bm{n} \cdot \partial_\mu \bm{\omega}') \} . 
\end{align}
This implies the transformation for the sum
\begin{align}
 \delta_\omega'( \mathbb{E}_{\mu\nu}+\mathbb{H}_{\mu\nu})
=& g (\mathbb{E}_{\mu\nu}+\mathbb{H}_{\mu\nu}) \times \bm{\omega}' ,
\end{align}
leading to the full SU(2)$_{\rm II}$ invariance:
\begin{align}
 \delta_\omega'( \mathbb{E}_{\mu\nu}+\mathbb{H}_{\mu\nu})^2 
=&  0 .
\end{align}
Moreover, we can show that 
\begin{align}
 \delta_\omega' ( D_\mu[\mathbb{V}] \mathbb{X}_\nu - D_\nu[\mathbb{V}] \mathbb{X}_\mu) 
=& g (D_\mu[\mathbb{V}] \mathbb{X}_\nu - D_\nu[\mathbb{V}] \mathbb{X}_\mu) \times \bm{\omega}' ,
\\
 \delta_\omega' (\mathbb{X}_\mu \times \mathbb{X}_\nu) =&   g (\mathbb{X}_\mu \times \mathbb{X}_\nu) \times \bm{\omega}' ,
\end{align}
which lead to the full SU(2)$_{\rm II}$ gauge invariance:
\begin{align}
 \delta_\omega' (D_\mu[\mathbb{V}] \mathbb{X}_\nu - D_\nu[\mathbb{V}] \mathbb{X}_\mu)^2 
=  0 ,
\quad 
 \delta_\omega' (\mathbb{X}_\mu \times \mathbb{X}_\nu)^2 =  0 ,
\\
  \delta_\omega'[(\mathbb{E}_{\mu\nu} + \mathbb{H}_{\mu\nu}) \cdot (g \mathbb{X}_\mu \times \mathbb{X}_\nu)] 
= 0. 
\end{align}

In particular, when $\bm{\omega}'(x)$ is parallel to $\bm{n}$, i.e., $\bm{\omega}'(x)=\theta'(x) \bm{n}(x)$, we obtain 

\noindent
\underline{Local U(1) gauge transformation II for $\bm{\omega}'(x)=\theta' (x) \bm{n}(x)$}:
\begin{subequations}
\begin{align}
  \delta_{\theta}' \bm{n}  =& 0  ,
\\
 \delta_{\theta}' c_\mu =&     \partial_\mu \theta'   ,
\\
  \delta_{\theta}' \mathbb{X}_\mu =&  g \mathbb{X}_\mu \times \theta'  \bm{n} ,
\\
\Longrightarrow  & \delta_{\theta}' \mathbb{B}_\mu  
=   0  ,
\quad
\delta_{\theta}' \mathbb{V}_\mu =    \bm{n} \partial_\mu \theta'    . 
\end{align}
\end{subequations}
Note that $\bm{n}$ and $\mathbb{B}_\mu$ are invariant under the U(1)$_{\rm II}$ gauge transformation II, while $c_\mu$ transforms as the U(1)$_{\rm II}$ gauge field.  
It is easy to show the local  U(1)$_{\rm II}$ gauge invariance for the field strengths:
\begin{align}
  \delta_{\theta}' \mathbb{E}_{\mu\nu}  =  0 , 
\quad
  \delta_{\theta}' \mathbb{H}_{\mu\nu} =  0 ,
\end{align}
which is also consistent with the initial definitions:
\begin{align}
  \mathbb{E}_{\mu\nu} = \bm{n}(\partial_\mu c_\nu - \partial_\nu c_\mu) ,
  \quad
  \mathbb{H}_{\mu\nu}  = - g \mathbb{B}_\mu \times \mathbb{B}_\nu .
\end{align}

Therefore, the dimension two composite operators $\mathbb{B}_\mu^2$, $\sqrt{\mathbb{H}_{\mu\nu}^2}$, and the dimension four operators  $\mathbb{H}_{\mu\nu}^2$ are gauge invariant under the local U(1)$_{\rm II}$ gauge transformation.

\subsection{MAG as a partial gauge fixing}

The gauge transformation I defined in the previous paper \cite{Kondo04} is nothing but $\delta_\omega$.
On the other hand, the gauge transformation II has been defined in  \cite{Kondo04} as a gauge transformation such that it does not change $\mathbb X^2$.  To see this, we consider the gauge transformation of $\mathbb X^2$.  
Since  the relationship (\ref{XnA}) leads to 
\begin{align}
   \mathbb X_\mu^2
 =  g^{-2} ({\bm n}\times D_\mu[\mathscr{A}]{\bm n})^2
=  g^{-2}
    \left\{
     (D_\mu[\mathscr{A}]{\bm n})^2
     -({\bm n}\cdot D_\mu[\mathscr{A}]{\bm n})^2
    \right\}
=  g^{-2} (D_\mu[\mathscr{A}]{\bm n})^2 ,
\end{align}
the gauge transformation of $\mathbb X^2$ is calculated as \cite{KMS05}
\begin{align}
\delta\frac12\mathbb X_\mu^2
  &=g^{-1}
     (D_\mu[\mathscr{A}]{\bm n}) \cdot 
     \{\bm n\times D_\mu[\mathscr{A}](\bm\theta^\perp-\bm\omega^\perp)\} ,
\label{eq:dX^2}
\end{align}
where we have used (\ref{gtn}) and (\ref{gtA}). 
Therefore, it turns out that the gauge transformation II corresponds to a special case 
$\bm\theta^\perp(x)=\bm\omega^\perp(x)$.

The average over the spacetime of  (\ref{eq:dX^2}) reads \cite{KMS05}
\begin{align}
\delta\int d^4x\frac12\mathbb X_\mu^2
  &=\int d^4x
     (\bm\theta^\perp-\bm\omega^\perp)\cdot
     D_\mu[\mathbb V]\mathbb X_\mu ,
\end{align}
where we have used (\ref{def:X})  
and integration by parts. 
Hence the minimizing condition 
\begin{align}
 0 = \delta\int d^4x\frac12\mathbb X_\mu^2 
 \label{MAGcond}
\end{align}
for arbitrary $\bm\theta^\perp$ and $\bm\omega^\perp$
yields the differential form:
\begin{equation}
\mathbb F_{\rm MA} = \bm{\chi} 
 :=D_\mu[\mathbb V]\mathbb X_\mu
 \equiv0 ,
\end{equation}
which reproduces exactly the MAG for the CFN variables \cite{Kondo04}. 
Therefore, the minimization condition (\ref{MAGcond}) works as a gauge fixing condition
except for the gauge transformation II, i.e., $\bm\theta^\perp(x)=\bm\omega^\perp(x)$.

\section{Lattice CFN variables and gauge fixing}
\setcounter{equation}{0}

\subsection{Continuum}

We show that for the CFN decomposition,
\begin{align}
  \mathscr{A}_\mu = \mathbb{V}_\mu + \mathbb{X}_\mu
= \mathbb{C}_\mu + \mathbb{B}_\mu + \mathbb{X}_\mu ,
\quad 
 \mathbb{C}_\mu = c_\mu \bm{n}, \quad 
 \mathbb{B}_\mu = g^{-1} \partial_\mu \bm{n} \times \bm{n} , 
\end{align}
the equality holds,
\begin{align}
  (D_\mu[\mathscr{A}] \bm{n})^2 = g^2 \mathbb{X}_\mu^2 .
\end{align}
In other words, $\mathbb{X}_\mu^2$ is rewritten in terms of $\mathscr{A}$ and $\bm{n}$. 
This is shown as follows.
\begin{align}
  (D_\mu[\mathscr{A}] \bm{n})^2
=& (\partial_\mu \bm{n} + g\mathscr{A}_\mu \times \bm{n})^2
\nonumber\\
=& (\partial_\mu \bm{n})^2  + 2g \partial_\mu \bm{n} \cdot (\mathscr{A}_\mu \times \bm{n}) + g^2 (\mathscr{A}_\mu \times \bm{n}) \cdot  (\mathscr{A}_\mu \times \bm{n})
\nonumber\\
=& (\partial_\mu \bm{n})^2  + 2g  \mathscr{A}_\mu \cdot (\bm{n} \times \partial_\mu \bm{n}) + g^2(\mathscr{A}_\mu \cdot \mathscr{A}_\mu)(\bm{n} \cdot \bm{n}) - g^2(\mathscr{A}_\mu \cdot \bm{n})^2 
\nonumber\\
=& g^2\mathbb{B}_\mu^2  - 2g^2  (\mathbb{C}_\mu + \mathbb{B}_\mu + \mathbb{X}_\mu) \cdot \mathbb{B}_\mu + g^2(\mathbb{C}_\mu + \mathbb{B}_\mu + \mathbb{X}_\mu)^2 - g^2(\mathbb{C}_\mu)^2 
\nonumber\\
=&  g^2\mathbb{B}_\mu^2  - 2g^2 \mathbb{B}_\mu^2 - 2g^2 \mathbb{X}_\mu \cdot \mathbb{B}_\mu + g^2(\mathbb{B}_\mu + \mathbb{X}_\mu)^2   
\nonumber\\
=& g^2\mathbb{X}_\mu^2 ,
\end{align}
where we have used
$
 (\mathbb{A} \times \mathbb{B}) \cdot (\mathbb{C} \times \mathbb{D}) = (\mathbb{A} \cdot \mathbb{C})(\mathbb{B} \cdot \mathbb{D}) - (\mathbb{A} \cdot \mathbb{D})(\mathbb{B} \cdot \mathbb{C}) 
$.

Another (simpler) way of showing the equivalence between $(D_\mu[\mathscr{A}] \bm{n})^2$ and $\mathbb{X}_\mu^2$ is as follows. 
By making use the fact that
\begin{align}
  D_\mu[\mathbb{V}] \bm{n} := \partial_\mu \bm{n} + g\mathbb{V}_\mu \times \bm{n} =  \partial_\mu \bm{n} + g\mathbb{B}_\mu \times \bm{n}
:= D_\mu[\mathbb{B}] \bm{n} \equiv 0 , 
\end{align}
we find
\begin{align}
  D_\mu[\mathscr{A}] \bm{n} = \partial_\mu \bm{n} +  g\mathscr{A}_\mu \times \bm{n} = \partial_\mu \bm{n} + g\mathbb{V}_\mu \times \bm{n} + g\mathbb{X}_\mu \times \bm{n} 
= g\mathbb{X}_\mu \times \bm{n} .
\end{align}
This fact leads us to the equivalence,
\begin{align}
  (D_\mu[\mathscr{A}] \bm{n})^2 = g^2(\mathbb{X}_\mu \times \bm{n})^2
= g^2(\mathbb{X}_\mu \cdot \mathbb{X}_\mu)(\bm{n} \cdot \bm{n}) - g^2(\mathbb{X}_\mu \cdot \bm{n})^2
= g^2\mathbb{X}_\mu^2 .
\end{align}
Hence, $\mathbb{X}_\mu$ is rewritten in terms of $\mathscr{A}_\mu$ and $\bm{n}$
\begin{align}
  \mathbb{X}_\mu = g^{-1} \bm{n} \times D_\mu[\mathscr{A}] \bm{n} .
\end{align}
This is also the case for $c_\mu$,
\begin{align}
  c_\mu =  \bm{n} \cdot \mathscr{A}_\mu .
\end{align}


As shown in Appendix A, we can impose the gauge fixing condition by minimizing the following functional 
 under the local gauge transformation:
\begin{align}
 0=  \delta \int_{x} \frac{1}{2} \mathbb{X}_\mu^2 ,
\end{align}
which leads to the differential form of the MAG condition in the CFN decomposition
\begin{align}
  D_\mu[\mathbb{V}] \mathbb{X}_\mu = 0 .
\end{align}
The MAG condition can also be derived by minimizing the functional
\begin{align}
 0=  \delta  \int_{x} \frac{1}{2} (D_\mu[\mathscr{A}] \bm{n})^2 
.
\end{align}
This is confirmed by explicit calculation and we leave it for the reader as an exercise.

\subsection{Lattice}

We show that the link variable on the lattice is identified with the CFN decomposition of the gauge potential as 
\begin{align}
  U_{x,\mu} = \exp \{ -i \epsilon g \mathscr{A}_\mu(x) \}
= \exp \{ -i \epsilon g[\mathbb{C}_\mu(x) + \mathbb{B}_\mu(x) + \mathbb{X}_\mu(x)] \} .
\label{U1}
\end{align}
In fact, we recover $(D_\mu[\mathscr{A}] \bm{n})^2$ in the naive continuum limit from the lattice functional
\begin{align}
 F_{MAG} := - \sum_{x,\mu} {\rm tr}[\bm{n}_x U_{x,\mu} \bm{n}_{x+\mu} U_{x,\mu}^\dagger ] .
\end{align}
In fact, by expanding the exponential $U_{x,\mu} = e^{ -i\epsilon \mathscr{A}_\mu(x) }$ into the Taylor series, we obtain
\begin{align}
& \frac14 \sum_{\mu} {\rm tr}[\bm{n}_x U_{x,\mu} \bm{n}_{x+\mu} U_{x,\mu}^\dagger ]
\nonumber\\
=& \frac{D}{2}- \frac{1}{4} \epsilon^2 \left\{ 
(\partial_\mu \bm{n})^2  - 2g   (\partial_\mu \bm{n} \times \bm{n}) \cdot \mathscr{A}_\mu  + g^2(\bm{n} \cdot \bm{n})(\mathscr{A}_\mu \cdot \mathscr{A}_\mu) - g^2(\bm{n} \cdot \mathscr{A}_\mu )^2
\right\} 
+ O(\epsilon^3) , 
\nonumber\\
=& \frac{D}{2}- \frac{1}{4} \epsilon^2  (D_\mu[\mathscr{A}] \bm{n})^2   
+ O(\epsilon^3)  
\nonumber\\
=& \frac{D}{2}- \frac{1}{4} \epsilon^2  g^2 \mathbb{X}_\mu^2  
+ O(\epsilon^3) , 
\label{F1c}
\end{align}
where the summation over $\mu=1, \cdots, D$ should be understood and the order $\epsilon$ terms cancel.
Therefore, we can obtain the MAG in the CFN decomposition by minimizing the functional $F_{MAG}$ with respect to the gauge transformation under the identification (\ref{U1}).

In particular, the naive MAG for the usual Cartan decomposition is obtained from minimizing the functional
\begin{align}
& \frac14 \sum_{\mu} {\rm tr}[\sigma_3 \ {}^GU_{x,\mu} \sigma_3 \ {}^GU_{x,\mu}^\dagger ]
\nonumber\\
=& \frac{D}{2}- \frac{1}{4} \epsilon^2 g^2 \left\{ 
 (\mathscr{A}_\mu \cdot \mathscr{A}_\mu) - (\sigma_3 \cdot \mathscr{A}_\mu )^2
\right\} 
+ O(\epsilon^3)  
\nonumber\\
=& \frac{D}{2}- \frac{1}{4} \epsilon^2  g^2 A_\mu^a A_\mu^a    
+ O(\epsilon^3) , 
\end{align}
for the link variable
\begin{align}
  {}^gU_{x,\mu} = \exp \{ \mp i \epsilon g \mathscr{A}_\mu(x) \} 
= \exp \{ \mp i \epsilon g[a_\mu(x) T^3 + A_\mu^a(x) T^a ] \} .
\end{align}

In the above calculation of the naive continuum limit, we have used the following formulae for the trace of the product of the generators in the SU(N) algebra.
\begin{align}
  {\rm tr}(T^A T^B) =& \frac{1}{2} \delta^{AB} ,
\\
  {\rm tr}(T^A T^B T^C) =& \frac{1}{4} (i f^{ABC} + d^{ABC}) ,
\\
  {\rm tr}(T^A T^B T^C T^D) =& \frac{1}{4N} \delta^{AB} \delta^{CD}
+ \frac{1}{8}(i f^{ABE} + d^{ABE})(i f^{CDE} + d^{CDE}) .  
\end{align}
They are obtained by the repeated use of 
\begin{align}
  T^A T^B = \frac{1}{2} [T^A, T^B] + \frac{1}{2} \{ T^A, T^B \}
= \frac{i}{2} f^{ABD} T^D + \frac{1}{2} \left( \frac{1}{N} \delta^{AB} + d^{ABD}T^D \right) .
\end{align}
 For SU(2), they are simplified as 
\begin{align}
  {\rm tr}(T^A T^B) =& \frac{1}{2} \delta^{AB} ,
\\
  {\rm tr}(T^A T^B T^C) =& \frac{1}{4}  i \epsilon^{ABC}   ,
\\
  {\rm tr}(T^A T^B T^C T^D) =& \frac{1}{8} \delta^{AB} \delta^{CD}
- \frac{1}{8} \epsilon^{ABE} \epsilon^{CDE}  
=  \frac{1}{8} \delta^{AB} \delta^{CD} 
- \frac{1}{8} \delta^{AC} \delta^{BD} + \frac{1}{8} \delta^{AD} \delta^{BC} .  
\end{align}
They are obtained by the repeated use of 
\begin{align}
  T^A T^B =  
  \frac{i}{2} \epsilon^{ABD} T^D +  \frac{1}{4} \delta^{AB}   .
\end{align}

In order to obtain this result (\ref{F1c}), we must symmetrize the expression, i.e.,
\begin{align}
  F_{MAG} :=& - \frac{1}{2} \sum_{x,\mu} {\rm tr}[\bm{n}_x U_{x,\mu} \bm{n}_{x+\mu} U_{x,\mu}^\dagger +\bm{n}_{x-\mu} U_{x-\mu,\mu} \bm{n}_{x} U_{x-\mu,\mu}^\dagger ]  
\nonumber\\
  =& - \frac{1}{2} \sum_{x,\mu} {\rm tr}[\bm{n}_x U_{x,\mu} \bm{n}_{x+\mu} U_{x,\mu}^\dagger +\bm{n}_{x} U_{x,-\mu} \bm{n}_{x-\mu} U_{x,-\mu}^\dagger ] ,
\end{align}
since $+\mu$ and $-\mu$ should be treated on the equal footing. 
Then the kinetic term for $\bm{n}$ is derived as
\begin{align}
  \frac{1}{2} \sum_{x,\mu} {\rm tr}[\bm{n}_x \bm{n}_{x+\mu} + \bm{n}_{x-\mu} \bm{n}_{x}]
=& \frac{1}{2} \sum_{x,\mu} {\rm tr}[\bm{n}_x (\bm{n}_{x+\mu} + \bm{n}_{x-\mu})]\nonumber\\
=& \frac{1}{2} \sum_{x,\mu} {\rm tr}[\bm{n}_x (2 \bm{n}_{x} +  \epsilon^2 \partial{}^{\epsilon}_\mu \partial'{}^{\epsilon}_\mu \bm{n}_{x})]
\nonumber\\
=& \frac{1}{2} \sum_{x} [D \bm{n}_{x} \cdot \bm{n}_{x} +  \frac{1}{2} \epsilon^2 \bm{n}_{x} \cdot \partial{}^{\epsilon}_\mu \partial'{}^{\epsilon}_\mu \bm{n}_{x} ]
\nonumber\\
=&  \sum_{x} \left[ \frac{D}{2}   -  \frac{1}{4} \epsilon^2  (\partial{}^{\epsilon}_\mu \bm{n}_{x})^2 \right]
,
\end{align}
where we have defined the forward derivative and the backward derivative by
\begin{align}
  \partial^{\epsilon}_\mu \phi(x) := [\phi(x+\mu)-\phi(x)]/\epsilon, \quad 
  \partial'{}^{\epsilon}_\mu \phi(x) := [\phi(x)-\phi(x-\mu)]/\epsilon , 
\end{align}
and the integration by parts,
\begin{align}
  \sum_{x} f(x) \partial'{}^{\epsilon}_\mu g(x) = - \sum_{x} \partial{}^{\epsilon}_\mu f(x) g(x) . 
\end{align}
The lattice d'Alembertian is defined by
\begin{align}
  \Delta(x,y) := - \partial{}^{\epsilon}_\mu \partial'{}^{\epsilon}_\mu  \delta(x-y)
= - \partial'{}^{\epsilon}_\mu \partial{}^{\epsilon}_\mu  \delta(x-y) .
\end{align}

The other terms can be calculated in the similar way. 

\subsection{Remarks}

We show that both $(D_\mu[\mathscr{A}] \bm{n})^2$ and $\mathbb{X}_\mu^2$ are invariant  
\begin{align}
  \delta_{II} (D_\mu[\mathscr{A}] \bm{n})^2 = g^2 \delta_{II} \mathbb{X}_\mu^2 = 0 ,
\end{align}
under the local gauge transformation II of the CFN variables:
\begin{align}
  \bm{n}(x) &\rightarrow U(x) \bm{n}(x) U^\dagger(x) := \bm{n}'(x) ,
\nonumber\\
  \mathbb{X}_\mu(x) &\rightarrow U(x) \mathbb{X}_\mu(x) U^\dagger(x) := \mathbb{X}_\mu'(x),
\nonumber\\
  D_\mu[\mathscr{A}](x) &\rightarrow U(x) D_\mu[\mathscr{A}](x) U^\dagger(x) := D_\mu[\mathscr{A}]'(x) ,
\end{align}
where $U(x)=\exp (ig\omega'(x))$. 
Note that $\bm{n}(x)$ and $\mathbb{X}_\mu(x)$ transform in the adjoint transformation under the gauge transformation II. 
In particular, $D_\mu[\mathscr{A}](x)$ transforms in the same way under the gauge transformation I and II, since it is written in terms of the original variable $\mathscr{A}_\mu$ which transforms as
\begin{align}
  \mathscr{A}_\mu(x) \rightarrow U(x) \mathscr{A}_\mu U^\dagger(x) + {i \over g} U(x) \partial_\mu U^\dagger(x) 
:= \mathscr{A}'_\mu(x) .
\end{align}

The signature in front of $\mathscr{A}$ is important, since 
\begin{align}
  (D_\mu[-\mathscr{A}] \bm{n})^2
=& (\partial_\mu \bm{n} - g\mathscr{A}_\mu \times \bm{n})^2
\nonumber\\
=& (\partial_\mu \bm{n})^2  - 2g \partial_\mu \bm{n} \cdot (\mathscr{A}_\mu \times \bm{n}) + g^2(\mathscr{A}_\mu \times \bm{n}) \cdot  (\mathscr{A}_\mu \times \bm{n})
\nonumber\\
=& (\partial_\mu \bm{n})^2  - 2g  \mathscr{A}_\mu \cdot (\bm{n} \times \partial_\mu \bm{n}) + g^2(\mathscr{A}_\mu \cdot \mathscr{A}_\mu)(\bm{n} \cdot \bm{n}) - g^2(\mathscr{A}_\mu \cdot \bm{n})^2 
\nonumber\\
=& g^2\mathbb{B}_\mu^2  + 2g^2  (\mathbb{C}_\mu + \mathbb{B}_\mu + \mathbb{X}_\mu) \cdot \mathbb{B}_\mu + g^2(\mathbb{C}_\mu + \mathbb{B}_\mu + \mathbb{X}_\mu)^2 - g^2(\mathbb{C}_\mu)^2 
\nonumber\\
=&  g^2\mathbb{B}_\mu^2  + 2g^2 \mathbb{B}_\mu^2 + 2g^2 \mathbb{X}_\mu \cdot \mathbb{B}_\mu + g^2(\mathbb{B}_\mu + \mathbb{X}_\mu)^2   
\nonumber\\
=& 4g^2\mathbb{B}_\mu^2 + 4g^2 \mathbb{X}_\mu \cdot \mathbb{B}_\mu + g^2\mathbb{X}_\mu^2
= g^2(2\mathbb{B}_\mu+\mathbb{X}_\mu)^2 ,
\end{align}
and there is no guarantee for the invariance of $(2\mathbb{B}_\mu+\mathbb{X}_\mu)^2$ under the gauge transformation II.

Note that in the usual continuum limit the signature in front of $\mathscr{A}_\mu$ is not important.  However, in this case, we can not adopt the 
\begin{align}
  U_{x,\mu} = \exp \{ i \epsilon g \mathscr{A}_\mu(x) \}
= \exp \{ i \epsilon g[\mathbb{C}_\mu(x) + \mathbb{B}_\mu(x) + \mathbb{X}_\mu(x)] \} .
\label{U2}
\end{align}
The identification (\ref{U1}) is necessary in order to recover the continuum expressions in the naive continuum limit as shown below. 
This is because the odd term in $\mathscr{A}_\mu$ plays the important role in this case and we can not change the signature arbitrarily. 

However, if we adopted (\ref{U2}), 
then the naive continuum limit left the $\mathbb{B}_\mu$ dependent term after the CFN decomposition, 
\begin{align}
& \frac14 \sum_{\mu} {\rm tr}[\bm{n}_x U_{x,\mu} \bm{n}_{x+\mu} U_{x,\mu}^\dagger ]
\nonumber\\
=& \frac{D}{2}- \frac{1}{4} \epsilon^2 \left\{ 
(\partial_\mu \bm{n})^2  + 2g   (\partial_\mu \bm{n} \times \bm{n}) \cdot \mathscr{A}_\mu  + g^2(\bm{n} \cdot \bm{n})(\mathscr{A}_\mu \cdot \mathscr{A}_\mu) - g^2(\bm{n} \cdot \mathscr{A}_\mu )^2
\right\} 
+ O(\epsilon^3) , 
\nonumber\\
=& \frac{D}{2}- \frac{1}{4} \epsilon^2  (D_\mu[\mathscr{-A}] \bm{n})^2   
+ O(\epsilon^3)  
\nonumber\\
=& \frac{D}{2}- \frac{1}{4} \epsilon^2  g^2 (2\mathbb{B}_\mu+\mathbb{X}_\mu)^2  
+ O(\epsilon^3) . 
\end{align}

Note that 
\begin{align}
& \frac14 \sum_{\mu} {\rm tr}[\bm{n}_x U_{x,\mu} \bm{n}_{x} U_{x,\mu}^\dagger ]
\nonumber\\
=& \frac{D}{2}- \frac{1}{4} \epsilon^2 g^2 \left\{ 
 (\mathscr{A}_\mu \cdot \mathscr{A}_\mu) - (\bm{n} \cdot \mathscr{A}_\mu )^2
\right\} 
+ O(\epsilon^3) , 
\nonumber\\
=& \frac{D}{2}- \frac{1}{4} \epsilon^2 g^2  (\mathbb{B}_\mu+\mathbb{X}_\mu)^2  
+ O(\epsilon^3) . 
\end{align}
If we wish to adopt (\ref{U2}) as the definition of the link variable, we must change the definition of $\mathbb{B}_\mu$ as
\begin{align}
  \mathbb{B}_\mu = g^{-1} \bm{n} \times \partial_\mu \bm{n} ,
\end{align}
which is determined from the condition of the covariant constant for the different covariant derivative,
\begin{align}
  D_\mu[\mathbb{V}] \bm{n} := \partial_\mu \bm{n} - g\mathbb{V}_\mu \times \bm{n} =  \partial_\mu \bm{n} - g\mathbb{B}_\mu \times \bm{n}
:= D_\mu[\mathbb{B}] \bm{n} \equiv 0 . 
\end{align}
Then we obtain the naive continuum limit, 
\begin{align}
& \frac14 \sum_{\mu} {\rm tr}[\bm{n}_x U_{x,\mu} \bm{n}_{x+\mu} U_{x,\mu}^\dagger ]
\nonumber\\
=& \frac{D}{2}- \frac{1}{4} \epsilon^2 \left\{ 
(\partial_\mu \bm{n})^2  + 2g   (\partial_\mu \bm{n} \times \bm{n}) \cdot \mathscr{A}_\mu  + g^2(\bm{n} \cdot \bm{n})(\mathscr{A}_\mu \cdot \mathscr{A}_\mu) - g^2(\bm{n} \cdot \mathscr{A}_\mu )^2
\right\} 
+ O(\epsilon^3) , 
\nonumber\\
=& \frac{D}{2}- \frac{1}{4} \epsilon^2  (\partial_\mu \bm{n} - g\mathscr{A}_\mu \times \bm{n})^2   
+ O(\epsilon^3)  
\nonumber\\
=& \frac{D}{2}- \frac{1}{4} \epsilon^2  g^2 \mathbb{X}_\mu^2  
+ O(\epsilon^3) . 
\end{align}



\end{document}